# Perception Gaps in Risk, Benefit, and Value Between Experts and Public Challenge Socially Accepted AI

A Preprint

**Philipp Brauner**
Communication Science
RWTH Aachen University
Aachen, 52074

**Felix Glawe**
Communication Science
RWTH Aachen University
Aachen, 52074

**Gian Luca Liehner**
Communication Science
RWTH Aachen University
Aachen, 52074

**Luisa Vervier**
Communication Science
RWTH Aachen University
Aachen, 52074

**Martina Ziefle**
Communication Science
RWTH Aachen University
Aachen, 52074

August 26, 2025

## Abstract

Artificial Intelligence (AI) is reshaping many societal domains, raising critical questions about its risks, benefits, and the potential misalignment between public and academic perspectives. This study examines how the general public (N=1110)—individuals who interact with or are impacted by AI technologies—and academic AI experts (N=119)—those elites shaping AI development—perceive AI's capabilities and impact across 71 scenarios. These scenarios span domains such as sustainability, healthcare, job performance, societal inequality, art, and warfare. Participants evaluated these scenarios across four dimensions using the psychometric model: likelihood, perceived risk and benefit, and overall value (or sentiment). The results suggest significant differences: experts consistently anticipate higher probabilities, perceive lower risks, report greater benefits, and express more positive sentiment toward AI compared to the non-experts. Moreover, both groups apply different weighting schemes: experts discount risk more heavily relative to benefit than non-experts. Visual mappings of these evaluations uncover areas convergent evaluations (e.g., AI performing medical diagnoses or criminal use) as well as tension points (e.g., decision of legal cases, political decision making), highlighting areas where communication and policy interventions may be needed. These findings underscore a critical translational challenge: if AI research and deployment are to align with societal priorities, the perception gap between developers and the public must be better understood and addressed. Our results provide an empirical foundation for value-sensitive AI governance and trust-building strategies across stakeholder groups.

**Keywords** Public Perception • AI Governance • Risk Perception • Societal integration of technology • AI Alignment • Expert-layperson comparison • Science and Technology Studies

# 1 Introduction

Although the origins of Artifical Intelligence (AI) date back decades (McCarthy et al. 2006; Hopfield 1982; Rumelhart, Hinton, and Williams 1986), recent advances in algorithms, computing power, and the availability of training data have led to transformative breakthroughs and increased funding (Deng et al. 2009; Lecun, Bengio, and Hinton 2015; Statista 2022). AI is becoming increasingly integrated across sectors such as education (Chen, Chen, and Lin 2020), healthcare (Amunts et al. 2024), journalism (Diakopoulos 2019), forestry and farming (Holzinger et al. 2024), as well as production and manufacturing (Brauner et al. 2022). It offers benefits such as convenience, efficiency, and innovation (Bouschery, Blazevic, and Piller 2023). The recent advances in large language models (LLMs) have begun to reshape both professional work practices and human communication. For example, (Geng et al. 2025) found that linguistic patterns characteristic of LLMs are increasingly present in academic articles. Also, (Brachman et al. 2024) reported that knowledge workers integrate LLMs into their workflows.

AI also raises concerns—such as privacy infringements (Lim and Shim 2022), job displacement (Acemoglu and Restrepo 2017), algorithmic bias Sadek, Calvo, and Mougenot (2024), and ethical challenges (Awad et al. 2018)—that affect individuals, organizations, and society as a whole. Consequently, expectations regarding AI remain divided: some view it as a revolutionary tool that will improve our lives (Brynjolfsson and McAfee 2014; Makridakis 2017), while others emphasize its ethical implications and potential risks (Cath 2018; Bostrom 2003).

Computers and algorithms are products of human design and are therefore shaped by developers' values, assumptions, and biases. Their effects and the decisions they inform are not value-neutral and may lead to unintended or harmful consequences—such as perpetuating inequality or reinforcing existing biases (Friedman and Nissenbaum 1996; Nissenbaum 2001; Sadek, Calvo, and Mougenot 2024). The "*AI alignment problem*" aims to counterbalance this by ensuring that artificial intelligence systems act in accordance with human values, intentions, and preferences (Hristova, Magee, and Soldatic 2024; Gabriel 2020). This involves designing AI that reliably understands, interprets, and follows human-aligned objectives, even as it becomes more advanced and autonomous. It underscores the importance of examining both the similarities and differences in AI perception between experts and developers—who shape these technologies—and the general public, who ultimately use or are affected by them.

In this article, we queried the opinions of both academic AI experts and the general public regarding AI's future and capabilities, using 71 short scenarios. Each scenario presents a different potential capability or impact that AI might have on individuals, companies, or society. Because individual perceptions of risk and benefit shape attitudes, usage intentions, and actual behavior across various domains Huang, Dai, and Xu (2020), we measured the expected likelihood of occurrence, perceived risks and benefits, and the overall value (or valence or sentiment) associated with these topics. We analyzed the responses for similarities, differences, and patterns, and provide visual maps of the findings. This study highlights both shared and divergent views on AI between AI experts and the public, aiming to support the development of a research agenda for AI's future, identify areas that may require stronger regulation, and underscore the need for accessible public education to enhance understanding of AI and its implications.

# 2 Related Work

This section reviews related work in this emerging field. First, we outline prior research on the perception of AI. Next, we present studies comparing how experts and the general public perceive AI and related technologies. Lastly, we identify the research gap and derive our research questions.

## 2.1 Public Perception of AI

While AI has existed for decades, the public release and rapid adoption of ChatGPT at the end of 2022 (Hu 2023) sparked unprecedented academic interest in its broader implications and how the public perceives AI across various tasks and domains. As research in this area continues to evolve, many findings remain preliminary, and a consolidated understanding of AI perception has yet to emerge. Accordingly, the following section offers a brief and necessarily selective overview of this rapidly developing field.

Public perception of AI is multifaceted and shaped by several factors, including media representation, personal experience, and broader societal context.

Media plays a significant role in shaping public opinion about AI. Fast and Horvitz (2017) conducted an extensive analysis of three decades of AI coverage in The New York Times, observing a marked rise in public interest after 2009. Their findings indicate that coverage has generally been more positive than negative, balancing optimism with concerns over control and ethical issues. In recent years, reporting has reflected heightened enthusiasm for AI's potential, particularly in fields like healthcare, mobility, and education.

News coverage often emphasizes the benefits of AI while downplaying potential risks, contributing to a perception of AI as superior to human capabilities and fostering the anthropomorphization of technology (Puzanova, Tertyshnikova, and Pavlova, n.d.). Cave, Coughlan, and Dihal (2019) examined AI narratives in the UK and identified four optimistic and four pessimistic themes. These narratives frequently evoke anxiety, with only two highlighting benefits over risks, such as AI's potential to make life easier.

Furthermore, sentiment analysis of WIRED articles reveals an increase in polarized views, with both positive and negative sentiments intensifying over time (Moriniello et al. 2024). Sanguinetti and Palomo (2024) investigated how news outlets portray AI and found that coverage often frames AI as something to be feared, depicting it as an autonomous and opaque entity beyond human control. Using an AI anxiety index, the study analyzed newspaper headlines before and after the launch of ChatGPT, reporting both increased coverage and heightened negative sentiment.

Perceived risks and benefits play a crucial role in shaping public attitudes toward AI. Surveys indicate that the public views AI as both a risk and an opportunity. Common concerns include privacy violations and cybersecurity threats (Brauner et al. 2023), while perceived benefits are often associated with applications in urban services and disaster management (Schwesig et al. 2023; Yigitcanlar, Degirmenci, and Inkinen 2022).

Neri and Cozman (2019) argue that experts significantly influence public perceptions of AI risks. Their public statements can amplify awareness of certain threats, such as existential risks, which may be rooted more in expert discourse than in actual incidents. Lee et al. (2024) found that individuals with higher education levels, greater political interest, and more knowledge about ChatGPT tend to perceive AI as more risky. This finding challenges the conventional "knowledge deficit" model, suggesting that negative perceptions may stem from a critical mindset that engages with AI technology more cautiously.

Trust in AI varies across contexts, demographic groups, and individual attitudes. For instance, people tend to trust AI more in personal lifestyle applications but remain more skeptical about its use by companies and governments (Yigitcanlar, Degirmenci, and Inkinen 2022). Willingness to engage with AI is shaped by the perceived balance of risks and benefits, which differs across domains such as healthcare, transportation, and media (Schwesig et al. 2023). In addition to variations in trust, there remains a significant gap in public understanding of AI, which can contribute to irrational fears and misinformed beliefs about control. Promoting AI literacy is therefore essential to enable informed decision-making and support responsible innovation (Brauner et al. 2023).

Public perception of AI also varies significantly based on local context, political ideology, and exposure to science news. For example, people in the United States generally expect more benefits than harms from AI, with a substantial portion supporting regulation to mitigate potential risks (Elsey and Moss 2023). However, existential risks are not a primary concern for most; instead, the public tends to worry more about tangible issues such as job displacement.

Further, public perception of AI can vary significantly based on local context, political ideology, and exposure to science news. For instance, the people from the US generally expects more benefits than harms from AI, with a significant portion supporting regulation to mitigate potential risks (Elsey and Moss 2023). However, existential risks are not a prominent concern for most people, who tend to worry more about concrete issues like job loss.

Sindermann et al. (2022) examined cross-cultural differences in AI attitudes among Chinese and German participants, linking fear of AI to neuroticism in both groups, while also highlighting cultural variations in AI acceptance and concern. Kelley et al. (2021) surveyed over 10,000 participants across eight countries (Australia, Canada, the United States, South Korea, France, Brazil, India, and Nigeria), finding that respondents in developed nations predominantly expressed worry and futuristic expectations, whereas those in developing countries showed greater enthusiasm for AI's potential. In particular, South Korea emphasized AI's practical usefulness and future applications, though widespread uncertainty about its broader societal impact persisted across all regions. In Taiwan, science news consumption and respect for scientific authority positively influence AI perceptions (Wen and Chen 2024). Interestingly, a recent large cross-cultural study building on Hofstede's cultural dimensions found that AI perception is rather shaped by individual differences than by cultural context (Wang 2025).

Public discourse on AI often oscillates between fear and inflated expectations—particularly regarding artificial general intelligence (AGI), which remains largely speculative and fictional at present (Jungherr 2023). A survey by Ipsos (2022) found that the general public frequently lacks a nuanced understanding of AI's technical capabilities and limitations. Similarly, Pew Research (2023) reported that only a small percentage of Americans could accurately identify AI in everyday scenarios, highlighting widespread confusion about its scope and functionality. The Alan Turing Institute (2023) likewise observed that public understanding of AI varies considerably depending on education level and context. Common concerns tend to focus on automation and robotics, especially in relation to employment and security. This limited awareness contributes to persistent misconceptions and overly simplistic views of AI's societal impact, ultimately hindering informed public discourse.

In summary, public perception of AI is multifaceted, shaped by media narratives, perceived risks and benefits, levels of trust, and cultural and contextual factors.

### 2.2 Similarities and Differences in Risk Perception between Experts and the Public

Risk perception often differs markedly between experts and the general public or novices across a range of domains and AI in particular. In this section, we first outline findings from outside the technology context, followed by research specifically focused on AI.

First, research has shown that health experts and the general public often differ in their assessments of health risks (Krewski et al. 2012). Experts typically perceive behavioral health risks (such as smoking and obesity) as more significant, while the public may prioritize other concerns. This discrepancy underscores the importance of effective risk communication strategies to better align public perception with expert evaluations.

Similarly, public perception of risks related to industrial production facilities tends to be more subjective and emotionally driven, in contrast to the more objective evaluations made by safety professionals (Botheju and Abeysinghe 2015). Such misalignments call for two-way communication approaches to address concerns proactively and prevent unnecessary escalation.

In another context, Siegrist et al. (2007) compared public and expert perceptions of nanotechnology using the psychometric model (that our research builds on), surveying 375 laypeople and 46 experts. Findings revealed that differences in risk perception extended beyond knowledge gaps, reflecting underlying value-based judgments. In the case of environmental hazards, such as nuclear waste, experts and laypeople also differ markedly in their perceived risks. Here, risk perception is shaped more by attitudes and moral values than by cognitive factors (Sjöberg 1998). Laypeople, who generally have less technical knowledge, tend to rely on intuitive and emotional reasoning, whereas experts base their assessments more on technical evidence and probabilistic analysis (Sjöberg 1998).

This difference often leads to divergent views on policy and regulation, with members of the public perceiving higher levels of risk than experts (Siegrist et al. 2007). The public prioritize ethical and societal implications, while experts focus primarily on scientific and technical risks.

In contrast, when it comes to natural hazards such as hurricanes and cyclones, there is often a considerable degree of alignment between expert risk assessments and public perceptions—particularly in high-risk areas (Peacock, Brody, and Highfield 2005; Sattar and Cheung 2019). However, public risk perception in these contexts can still be shaped by factors such as trust in authorities and prior experience with disasters. In the domain of autonomous vehicles (AVs), public risk perception is strongly influenced by trust in both the technology and the institutions that regulate it. Greater knowledge about AVs can enhance trust, which in turn reduces perceived risk, highlighting the importance of targeted trust-building initiatives (Robinson-Tay and Peng 2024). In aviation, by contrast, experts typically possess a more accurate understanding of relative risks, while novices' perceptions may be distorted by overconfidence or limited experience (Thomson et al. 2004).

Elena and Johnson (2015) examined differences in expert and public perceptions of cloud computing services. The findings indicate that experts tend to have a more nuanced understanding of risks, particularly regarding data security and integrity, while members of the public are more likely to experience a generalized dread risk in response to unfamiliar or abstract technological threats. Perceptions were also influenced by factors such as trust in regulatory bodies and the perceived benefits of the technology.

A decade ago, Müller and Bostrom (2016) surveyed AI experts on their expectations for the future capabilities of artificial intelligence, finding that most anticipated the emergence of superintelligence between 2040 and 2050. Notably, one-third of these experts considered this development *bad* or *extremely bad*, highlighting substantial concerns even within the expert community.

Crockett et al. (2020) compared trust and risk perceptions of AI between the general public and computer science students as individuals with—supposedly—above-average expertise in AI. The findings revealed clear differences, suggesting that education plays a crucial role in increasing trust and reducing perceived risks. Similarly, Novozhilova et al. (2024) found that greater technological competence and familiarity with AI are associated with higher levels of trust in AI systems.

Still, comprehensive comparisons between experts and non-experts remain relatively rare. Recently, Jensen et al. (2024) conducted interviews with 25 members of the general public and 20 AI experts in the United States to examine their perceptions of AI. Both groups emphasized that AI systems reflect the values and biases of their creators, acknowledging inherent limitations in the technology. Ethical concerns cited by participants included AI's lack of transparency, its profit-driven development, and the risk of exacerbating existing social inequalities. Human oversight was widely supported, particularly in high-stakes contexts such as healthcare. Although AI is perceived as efficient, its inability to replicate human empathy emerged as a central barrier to trust. Across both groups, reflections on humanness and ethics played a critical role in shaping attitudes toward AI.

### 2.3 Research Questions

Integrating both expert and public perspectives is critical for shaping AI policies and ensuring alignment with human norms and values. Divergent perceptions of risk can significantly influence technology design, public acceptance, and regulatory outcomes. Although the literature on AI perception and use is growing, there remains a lack of comprehensive understanding of how expert and public views align or diverge regarding AI and its potential applications.

This study addresses these gaps through the following research questions:

1. Perception and Impact: How do AI experts (who shape and design AI-based systems) and members of the general public (who use or are affected by these systems) differ in their perceptions of AI's capabilities and impacts?
2. Value Attribution: What values do experts and the public assign to potential AI futures? In which domains is AI perceived positively or negatively, and how do these perceptions differ between the two groups?
3. Risks and Benefits: What risks and benefits are perceived in AI's potential futures? How do these perceptions differ between experts and the public?
4. Risk–Benefit Trade-offs: Is the overall value attributed to AI more strongly influenced by perceived risks or by perceived benefits? How do these trade-offs vary between experts and the public?

## 3 Method

The goal of this study is to identify similarities and differences in how the general public and academic AI experts perceive AI, with a focus on the underlying trade-offs between perceived risks, benefits, and overall value.

### 3.1 Risk Perception and the Psychometric model

This work builds on the psychometric model of risk perception, which focuses on individuals' subjective evaluations of risk, often measured through rating scales (Fischhoff et al. 1978; Slovic, Fischhoff, and Lichtenstein 1986). This model is well-suited to studying risk perception in emerging technologies and offers a structured framework for understanding how people evaluate and balance risks and benefits. The model has been applied in various contexts, including gene technology (Connor and Siegrist 2010), genetically modified food (Verdurme and Viaene 2003), nuclear energy (Slovic et al. 2000), climate change (Pidgeon and Fischhoff 2011), and carbon capture and utilization technologies (Arning et al. 2020).

A consistent pattern across these domains is the inverse relationship between perceived risk and perceived benefit: technologies considered more useful are often seen as safer, whereas those viewed as risky tend to be rated as less beneficial (Alhakami and Slovic 1994; Efendić et al. 2021).

### 3.2 Survey Design

To address our research questions, we designed two closely related surveys: one for the general public and another for academic AI experts. Both surveys centered on a shared core, in which participants evaluated a randomized random subset of 15 out of 71 micro-scenarios (to reduce participant fatigue), each describing a potential AI development or an imanginary AI future (Brauner 2024). Participants rated each scenario using five assessment items This design supports two complementary perspectives in interpreting the data: 1) Individual-level interpretation: responses serve as reflexive indicators of participants' underlying dispositions or differences. 2) Scenario-level interpretation: responses reflect how participants attribute expectation, risk, benefit, and value to specific technologies or topics, enabling pattern analysis and visual mapping across groups.

To develop the list of scenarios, we first identified potential capabilities and impacts that AI might have in the near future. Drawing on expert workshops and prior research (Brauner et al. 2023), we compiled an initial set of potential survey topics. For topic selection, we built on Luhmann's system theory ("Systemtheorie"), which conceptualizes modern society as comprising six core subsystems: economy, law, science, politics, religion, and education (Luhmann 1989). Each of these subsystems fulfills a unique and non-substitutable function within society (Peterson, Peterson, and Grant 2004).

Through multiple iterative rounds, we ensured that each subsystem was represented, removed redundant items, and refined the phrasing of topics for clarity and conciseness. The final set includes both plausible and more speculative AI-related developments, such as: "AI creates many jobs", "AI promotes innovation", "AI acts according to moral concepts", and "AI considers humans as a threat". The complete list of the 71 items used in the study is provided in Table 4 in the Appendix.

We selected five dependent variables to evaluate each topic, using single-item 6-point semantic differential scales. Single-item measures allow for efficient data collection and are considered appropriate for capturing well-defined constructs such as risk, benefit, and value judgments (Rammstedt and Beierlein 2014; Fuchs and Diamantopoulos 2009). The use of 6-point scales avoids midpoint bias by eliminating a neutral option, thereby encouraging more decisive responses (Tourangeau, Rips, and Rasinski 2000).

- Expected Likelihood: How likely is it that this development will occur within the next 10 years ("will not happen—will happen")?
- Personal Risk: How do you assess the risk of this development for yourself personally ("low risk—high risk")?
- Social Risk: How do you assess the risk of this development for society as a whole ("socially harmless—socially harmful")[1]?
- Perceived Benefit: How beneficial or useful do you consider this development ("useless—useful")?
- Attributed Value: Assuming this development comes true, would you consider it positive or negative ("negative—positive")?

To assess participants' overall evaluation of each topic, the final item drew on the Value-Based Adoption Model (VAM), which conceptualizes valence or attributed value—ranging from positive to negative—as a suitable criterion for technology evaluation (Kim, Chan, and Gupta 2007).

In addition to these topic-specific evaluations, demographic information was collected. All participants were asked to report their age (in *years*) and gender (following Spiel, Haimson, and Lottridge (2019) as *male*, *female*, *diverse*, or *prefer not to say*).

In the AI expert survey, participants also self-reported their level of expertise in AI using a five-point scale (*no expertise in the field of AI*, *basic knowledge*, *well-informed*, *expert*, *recognized authority in the field of AI*). Additional items assessed years of experience, number of scientific publications, current country of residence, and scientific subdiscipline.

In the general public survey, we collected data on education level, employment status, and several individual difference measures potentially influencing AI perception. These included Technology Readiness (Neyer, Felber, and Gebhardt 2016), an AI Readiness Scale (Karaca, Çalışkan, and Demir 2021), and interpersonal trust (Niessen et al. 2021). A detailed analysis of the impact of these individual factors is presented in a separate publication (Author and Writer 2024). Figure 1 illustrates the design of both surveys.

Both surveys began with an informed consent procedure, informing participants that their participation was voluntary, that no personally identifiable information would be collected, and that the resulting data would be made available as open data. The expert questionnaire was administered in English, while the public survey was conducted in German. The study received ethical approval from the university's Institutional Review Board (IRB) under the protocol ID . All materials, including the full survey design, data, and analysis scripts, are openly available in the OSF repository: https://osf.io/gt9un/?view_only=8963c39ca00f4b02a6800c218e222756.

### 3.3 Sample Acquisition

The public sample was acquired in collaboration with a research panel provider. A total of 1354 participants were recruited to approximate representativeness of the German population in terms of age, gender, socio-economic status, and regional distribution across the federal states. Participants received a monetary compensation of approximately 1 Euro for completing the survey.

[1]This item showed very strong correlations with personal risk across all topics and both samples, and was therefore excluded from further analysis. Researchers with a specific interest in social risk are encouraged to explore this variable independently. The data is publically available.

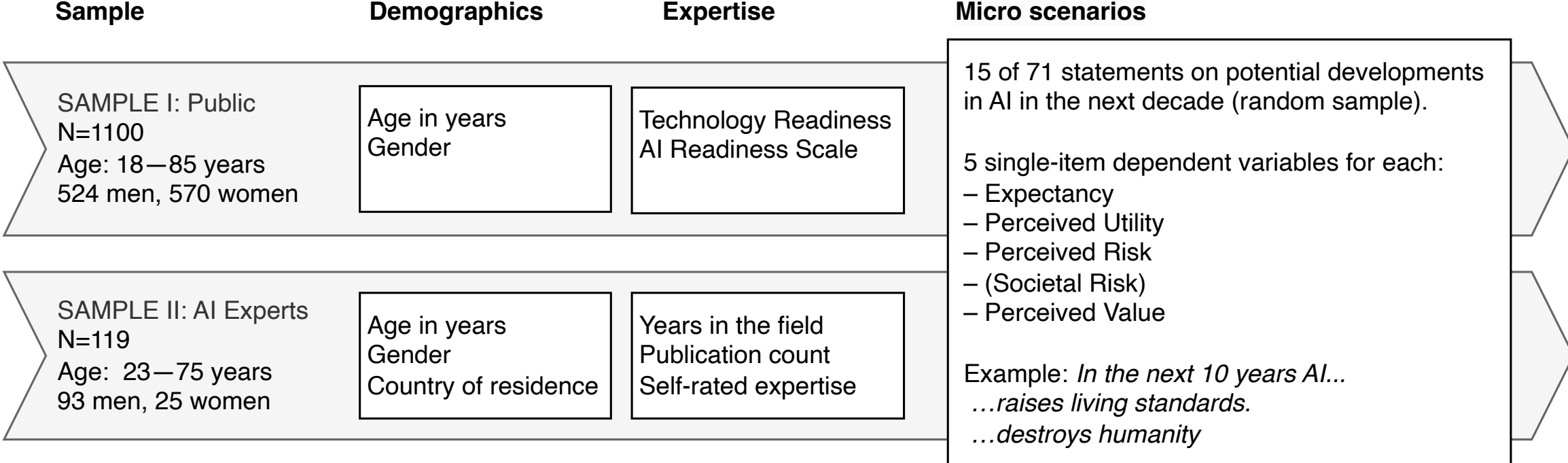

Figure 1: Overview of the experimental design for both surveys. In each survey, participants evaluated 15 randomly selected micro-scenarios from a pool of 71. The measured explanatory user factors differed between experts and the members of the public.

The expert sample was obtained via convenience and snowball sampling. We initially contacted individuals within our academic networks, including collaborators from joint projects and funding proposals. Additionally, we compiled a list of publicly available email addresses from authors of recent publications and conference presenters in AI-related venues. These contacts were invited via email and encouraged to forward the invitation to colleagues. In total, we contacted approximately 450 AI researchers worldwide, of which 139 proceeded beyond the landing page of the survey. To encourage participation, the expert invitation emphasized the academic relevance of the study and the open availability of its findings.

### 3.4 Data Cleaning and Analysis

All materials, data, analysis scripts, and this reproducible manuscript are publicly available in the open data repository. The analyses were conducted using R version 4.3.2 (2023-10-31). We employed both parametric and non-parametric statistical methods, including Bravais-Pearson ($r$) and Kendall's Tau ($\tau$) correlation coefficients, Chi-square ($\chi^2$), Ordinary least squares (OLS) multiple linear regressions with robust standard errors, and Multivariate analyses of variance (MANOVA) using Pillai's trace ($V$) as the omnibus test statistic. Due to substantial different group sizes between the expert's and public's samples, we ensured that all methods used are robust to sample imbalance. While OLS point estimates are generally reliable under unequal sample sizes (Fox 2016), standard errors may be biased. Hence, we calculated robust standard errors and further cross-validated key findings through random subsampling with equal group sizes, which yielded similar results. Consistent with social science standards, we set the Type I error rate to $\alpha = .05$ to determine statistical significance (Field 2009). We assessed the assumptions for each statistical method and report any violations where applicable. Missing responses were handled on a test-wise deletion basis.

We filtered the data from both samples for incomplete or low-quality responses using both reactive and non-reactive criteria (Leiner 2019). For the general public, participants were included only if they completed the survey and passed an attention-check item ("Please select 'rather agree'"). For the expert sample, we included only those who reported at least basic knowledge of AI and had more than one year of experience working in the field. In both samples, we excluded participants who completed the survey in less than one-third of the median completion time ("speeders"). This threshold is generally considered sufficient to identify invalid or inattentive responses (Leiner 2019). The median completion times of the surveys was 9.8 minutes for the laypoeple and 9.5 minutes for the experts.

Based on these criteria, we excluded 254 out of the 1354 participants in the general public sample (exclusion rate of 18.8%) and 20 of the 139 participants in the expert sample (exclusion rate of 14.4%). The unfiltered datasets and the filtering procedures are fully documented and available in the open data repository.

### 3.5 Sample 1: General Public

After cleaning, the sample of the general public consists of 1100 participants, with 524 identifying as male and 570 as female. The participants' ages ranged from 18—85 years, with a median age of 51 years (SD = 14.2) years. There was no significant correlation between age and gender in the sample ($\tau = -0.031$, $p = 0.210 > .05$).

### 3.6 Sample 2: Artificial Intelligence Experts

After cleaning, the expert sample consists of 119 participants with 93 identifying as male and 25 as female. Participants' ages ranged from 23—75 years, with a median age of 36.5 years (SD = 13.4 years). Again, there was no significant

Table 1: Comparison of average assessments across the four evaluation dimensions—expectancy, risk, benefit, and value—for AI experts and the general public. A MANOVA revealed a statistically significant overall group effect, as well as significant differences on each individual dependent variabl.

| | Experts | | | | Public | | | | Sig. |
|---|---|---|---|---|---|---|---|---|---|
| Dimension | M | SD | CI-95 | N | M | SD | CI-95 | N | p |
| Expectancy | 25.2% | 67.4% | [13.1%,37.3%] | 119 | 12.7% | 66.7% | [8.8%,16.7%] | 1100 | <0.001 |
| Perceived Risk | 19.3% | 63.1% | [7.9%,30.6%] | 119 | 34.7% | 56.4% | [31.4%,38.1%] | 1100 | <0.001 |
| Perceived Benefit | 15.3% | 62.1% | [4.1%,26.4%] | 119 | -5.2% | 59.0% | [-8.7%,-1.7%] | 1100 | <0.001 |
| Attributed Value | -4.0% | 60.9% | [-14.9%,7.0%] | 119 | -19.7% | 57.8% | [-23.2%,-16.3%] | 1100 | <0.001 |

*Note:*
Measured on 6-point Likert scales and rescaled to -100% to +100%. Negative values indicate a negative evaluation of the respective dimension (i.e., low value, low perceived risk, low perceived benefits, or low expectancy) and positive values indicate a high evaluation.

association between age and gender ($\tau = -0.113$, $p = 0.143 > .05$). The majority of the experts were from Germany (N = 77), followed by the Netherlands (N = 10). All other countries received less than 10 mentions.

Years of AI experience ranged from 1 to 40 years with an arithmetic mean of 10.3 and a median of 5 years. In total, the participants reported 2713 academic publications, with an arithmetic mean of 22.8 and a median of 5 publications per expert. As indicated by the high Gini coefficient ($G = 0.771$), the number of publications is unevenly distributed among the experts—many have few, while a few have many. This suggests that both prominent figures and junior researchers participated in the survey. When asked for their level of expertise, 11 participants reported having basic knowledge, 51 reported being well-informed, 48 reported being experts, and 9 described themselves as a "*recognized authority in the field of AI*".

## 4 Results

We begin by analyzing differences between the general public and academic AI experts in terms of *overall perceived risk*, *benefit*, *attributed value*, and expected *likelihood*, that is, how likely each of the many AI projections is to come true within the next decade. Next, we interpret participants' responses as reflexive measurements of latent constructs: expectancy, perceived risk, perceived benefit, and overall value of AI. These are treated as *individual-level personality differences*, allowing us to examine how individuals trade off perceived risks and benefits differently across the two groups. Finally, we shift the perspective and interpret responses at the *topic level*, focusing on how specific AI-related statements were evaluated. We assess how experts and the public differ in their expectations and valuations of these projections, including how they attribute risks and benefits to individual AI scenarios. Table 4 provides a complete overview of all ratings across all statements for both groups.

### 4.1 Do overall evaluation scores differ between academic AI experts and the general public?

In this section, we examine how both samples evaluated the AI-related topics across the queried dependent variables: perceived risk, benefit, attributed value, and expectancy (i.e., the expected likelihood that the respective AI development will occur within the next ten years). As illustrated in the box plots in Figure 2, notable differences emerge between the two groups. A one-way MANOVA with group (public vs. experts) as the independent variable revealed a small but statistically significant difference between the two samples on the combined dependent variables (V=0.057, F(4, 1214) =18.332, p<0.001).

Follow-up univariate analyses showed that each of the four dependent variables exhibited significant group differences: Expectancy: F(1, 1217) = 18.572, p<0.001, risk: F(1, 1217) = 26.716, p<0.001, benefit: F(1, 1217) = 43.443, p<0.001, and value: F(1, 1217) = 21.929, p<0.001). These findings indicate that academic AI expertise significantly influences not only the overall multivariate response but also each individual evaluation dimension.

In particular, AI experts, on average, rated the queried projections as more likely to materialize within the next decade (25.2%), less risky (19.3%), more useful (15.3%), and more positive overall (-4.0%) compared to the members of the general public ( expectancy: 12.7%, risk: 34.7%, benefit: -5.2%, value: -19.7%). Table 1 summarizes these findings.

Table 2: Correlation analysis of the assessment dimensions across the 71 topics for experts and the public.

| | | AI Experts (N=119) | | | Public (N=1100) | | |
|---|---|---|---|---|---|---|---|
| Variable 1 | Variable 2 | r | 95%-CI | p | r | 95%-CI | p |
| Expectancy | Risk | 0.144 | [-0.04,0.32] | 0.118 | 0.212 | [0.15,0.27] | <0.001 |
| Expectancy | Benefit | 0.401 | [0.24,0.54] | <0.001 | 0.143 | [0.08,0.20] | <0.001 |
| Expectancy | Value | 0.393 | [0.23,0.54] | <0.001 | 0.039 | [-0.02,0.10] | 0.193 |
| Risk | Benefit | -0.295 | [-0.45,-0.12] | 0.002 | -0.639 | [-0.67,-0.60] | <0.001 |
| Risk | Value | -0.418 | [-0.56,-0.26] | <0.001 | -0.749 | [-0.77,-0.72] | <0.001 |
| Benefit | Value | 0.712 | [0.61,0.79] | <0.001 | 0.869 | [0.85,0.88] | <0.001 |

*Note:*
Correlations of the average evaluations of the 71 queried topics (perspective 1).

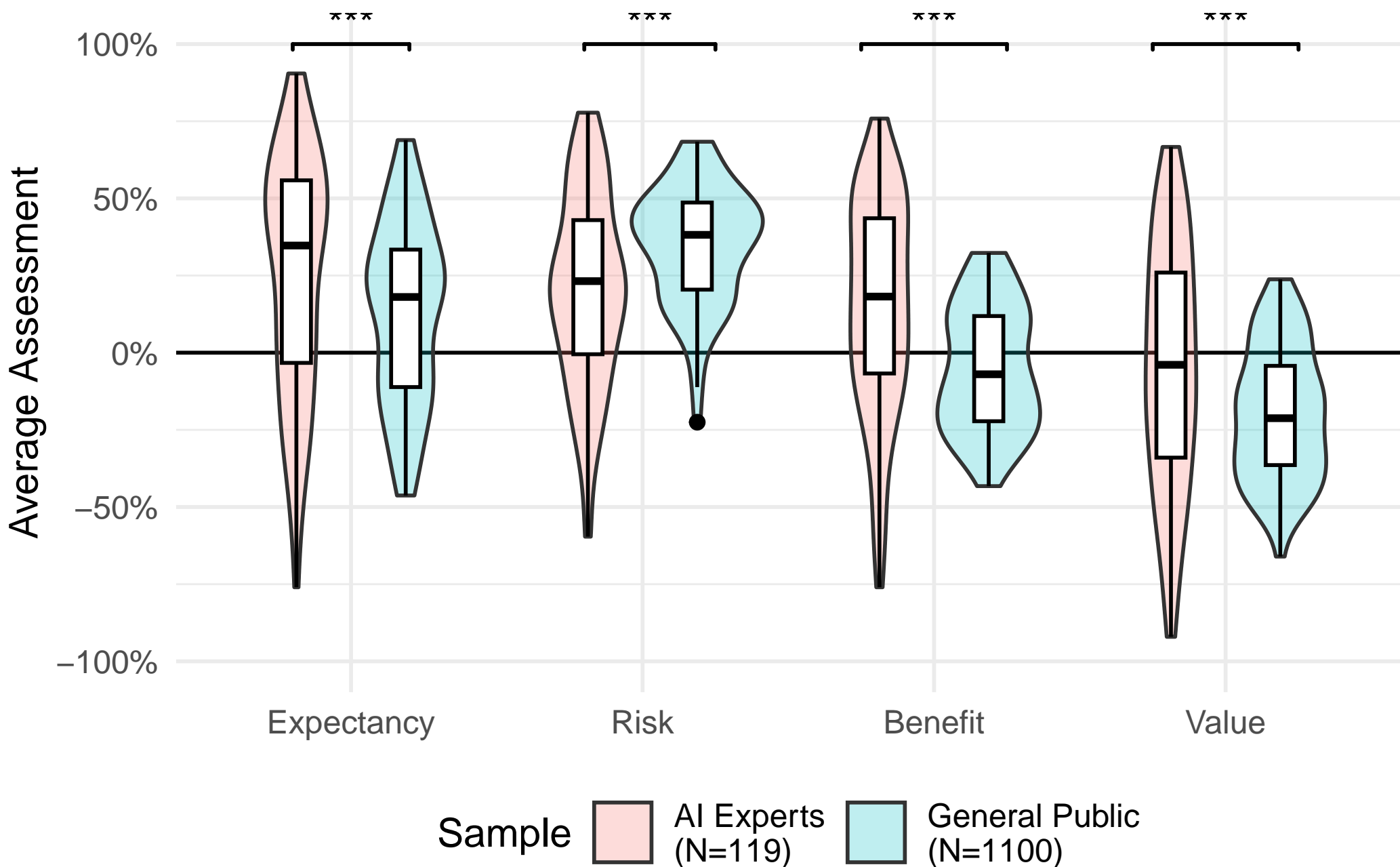

Figure 2: Box plots showing the overall average evaluations across all 71 topics for each assessment dimension, separately by sample (experts vs. public). The shaded areas represent the distribution density of the data. Asterisks indicate statistically significant differences between the two groups for each dimension.

### 4.2 Do Academic AI Experts and the Public Differ in Weighing AI Risks and Benefits?

In this section, we examine how individuals from both samples perceive AI in terms of expectancy, perceived risk and benefit, and overall attributed value. Hereto, we interpret the participants responses as reflexive measurements of latent psychological constructs.

Table 2 presents the correlation patterns between individuals' expectations of AI outcomes, perceived risks and benefits, and their overall value judgments of AI for both groups.

Among the academic AI experts, the expected likelihood of occurrence is positively associated with both perceived benefits and overall value. However, there is no significant relationship between expectancy and perceived risk. That is, experts' belief that AI projections will materialize is independent of their perception of associated risks. As expected, perceived risk and benefit are moderately negatively correlated. Perceived risk is also negatively associated with attributed value, while perceived benefit shows a strong positive relationship with overall value.

In contrast, the pattern is somewhat different among the general public. Expectancy is positively correlated with *both* perceived risk and perceived benefit, suggesting that individuals who see AI as either risky or beneficial are more likely to believe that the projections will come true in the near future. However, expectancy does not significantly correlate with overall value. The overall attributed value of AI is, as with the experts, strongly positively associated with perceived benefits and strongly negatively associated with perceived risks.

In both samples, the general public and AI experts, the overall perceived value of AI is negatively associated with perceived risks and positively associated with perceived benefits. Moreover, individuals' perceptions of AI risks and benefits are mutually correlated.

To disentangle these overlapping influences and assess the relative contribution of each predictor in the risk–benefit trade-off, we conducted three multiple linear regression analyses. First, we ran separate regressions for each sample, using perceived risks and perceived benefits as independent variables and the overall attributed value of AI as the dependent variable. Second, we conducted a combined-sample regression, including a sample identifier as a third predictor to assess group-level effects[2]. Table 3 summarizes the results of all three significant regression models.

The third regression model (right side of the table), which includes the factor distinguishing between the two samples, was significant and explained 80% of the variance in overall value (F(3,1215)=1773.106, $p < 0.001$, $R^2 = 0.807$). This group factor was significant, indicating a small but meaningful difference in the risk–benefit trade-offs between academic AI experts and the general public ($\beta = 0.044$, $p < 0.006$). Given these significantly different trade-offs, we proceed by analyzing the two samples separately in the following sections. Figure 3 illustrates both models.

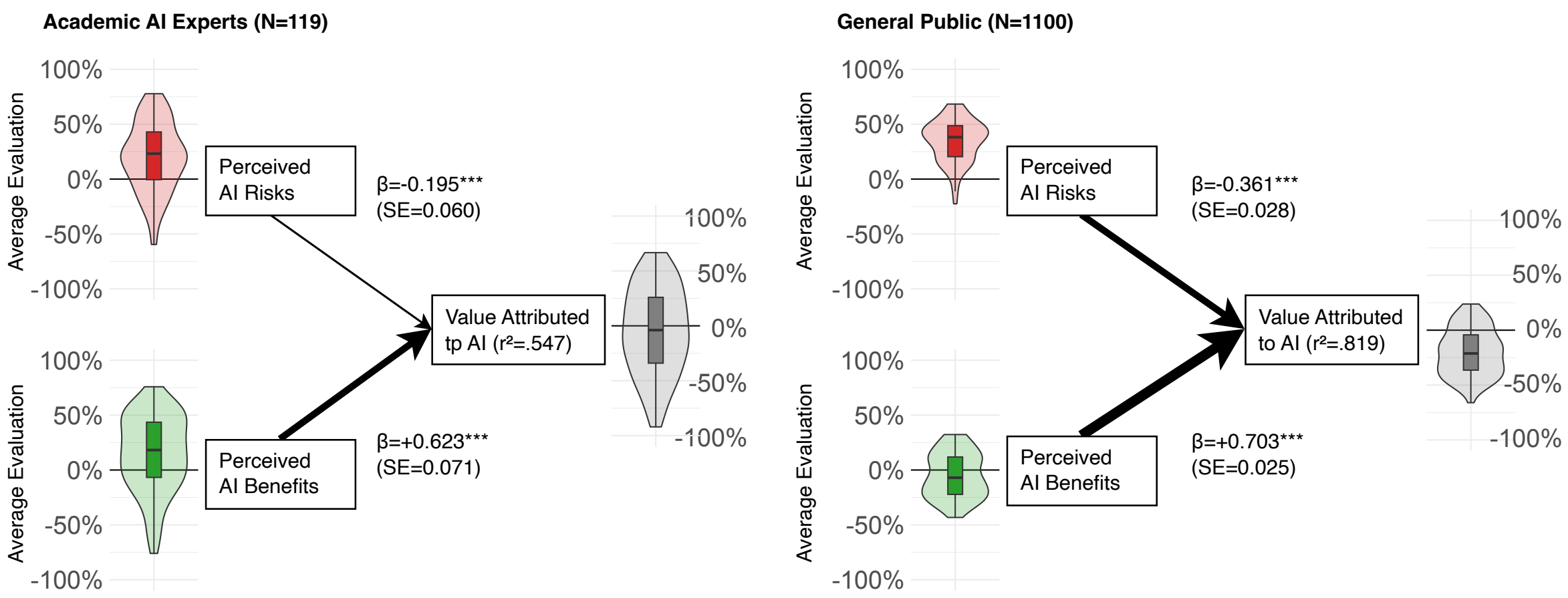

Figure 3: Visualization of the two significantly different multiple linear regression models for academic AI experts (left) and the general public (right), including variable distributions shown as violin and boxplots. In both models, perceived risk and benefit significantly predict attributed AI value. However, the explained variance and the influence of risk are greater in the public sample. Line thickness reflects the strength of each predictor's standardized effect ($\beta$). All predictors are significant at $p < .001$.

For the sample of academic AI experts (left side of the table), the regression model was significant, explaining over 54% of the variance in overall AI value ($F(2, 116) = 48.5$, $p < 0.001$, $R^2 = 0.547$. The intercept was significant and negative ($I = -0.101$), indicating a slightly negative baseline evaluation, i.e., if an expert held neutral views on both AI risks and benefits, their overall perceived value of AI would tend to be slightly negative. Perceived AI risk had a moderate negative effect on value ($\beta = -0.195$), whereas perceived AI benefit had a strong positive effect ($\beta = 0.623$). Importantly, the influence of perceived benefits was approximately three times stronger than the influence of perceived risk ($\times 3.19$).

For the general public (middle part of the table), the regression model was also significant and showed an even stronger fit, explaining 82% of the variance in overall AI value ($F(2, 1097) = 2634.6$, $p < 0.001$, $R^2 = 0.819$). In contrast to the expert group, the intercept was not significantly different from zero, indicating no consistent baseline tendency in AI evaluations when perceived risks and benefits are neutral. Overall value judgments were significantly influenced by both perceived risk ($\beta = -0.361$) and perceived benefit ($\beta = 0.703$) with benefits having about twice the impact of risks ($\times 1.95$). Notably, the negative effect of perceived risk was substantially and significantly larger for the general public compared to the expert group, highlighting a greater sensitivity to potential downsides of AI among lay respondents.

[2]While unequal group sizes can result in heteroskedasticity that affects the standard errors, this does not bias the regression coefficient estimates ($\beta$s) (Fox 2016). As cross-validation, we additionally performed a subsampling analysis using small random subsets of the general public sample ($N' = 110$), which yielded comparable results.

Table 3: Regression results from three multiple linear models examining the relationship between perceived AI risks and benefits (as rated by academic AI experts and the general public) and overall attributed value. The third model includes sample group as an additional predictor (combined sample). All models use robust standard errors to account for unequal sample sizes. Results indicate significant differences in risk–benefit trade-offs between experts and the public, with experts placing comparatively less emphasis on AI-related risks.

| | Dependent variable: Overall Attributed Value | | |
|---|---|---|---|
| Independent variables | Academic AI Experts (N=119) | General Public (N=1100) | Combined |
| (Intercept) | -0.101*** | -0.036 | -0.085*** |
| (SE) | (0.024) | (0.009) | (0.017) |
| Perceived Risk $\beta$ | -0.195*** | -0.361*** | -0.346*** |
| (SE) | (0.060) | (0.028) | (0.025) |
| Perceived Benefit $\beta$ | 0.623*** | 0.703*** | 0.704*** |
| (SE) | (0.071) | (0.025) | (0.023) |
| Sample $\beta$ | — | — | 0.044* |
| (SE) | | | (0.016) |
| $F$ | F(2,116)=48.526 | F(2,1097)=2634.605 | F(3,1215)=1773.106 |
| $p$ | <0.001 | <0.001 | <0.001 |
| adj. $R^2$ | 0.547 | 0.819 | 0.807 |

### 4.3 Do Experts and the Public Share the Same Vision for AI's Future?

As the previous analysis shows, academic AI experts and the general public differ in their expectations of what AI is likely to achieve over the next 10 years (hereafter referred to as expectancy). To illustrate these differences at the topic level, Figure 4 presents a scatter plot of the average expectancy ratings for each item.

Each point in the figure represents a single survey item, with its position determined by the mean expectancy rating from experts (on the $x$-axis) and from the public (on the $y$-axis). The plot serves as a comparative map of perceived likelihoods for both groups: points farther to the right indicate higher expectations among experts, while points higher on the plot reflect higher expectations among members of the public. Items that lie on or near the dashed diagonal indicate agreement between the two groups: experts and the public rated the likelihood of these scenarios similarly. In contrast, points that deviate substantially from the diagonal reflect disagreement. Items positioned above the diagonal are considered more likely by the public than by experts, while those below the diagonal are judged more likely by experts. The blue regression line indicates the best linear fit (maximum likelihood estimation), and the shaded gray band around it shows the 95% confidence interval, highlighting the uncertainty range for the estimated relationship.

The figure illustrates both notable similarities and distinct differences in expectancy assessments between the public and academic AI experts. First, there is an overall strong and positive correlation between the two groups' expectations regarding AI, indicating substantial agreement on the perceived likelihood of many AI-related developments ($r = 0.734$, $p < 0.001$). Second, the distribution of expectancy ratings among the public is markedly narrower than that of the experts, suggesting that experts hold a more nuanced and differentiated view of AI's future potential. This difference in variance is also reflected in the flatter slope of the regression line.

Third, despite this general agreement, several topics exhibit strong divergences between the two groups. Specifically, the public expressed higher expectations than experts for statements such as "AI will destroy humanity", "AI will lead to personal loneliness", and "AI can no longer be controlled by humans". Conversely, experts reported substantially higher expectations for statements like "AI will improve our health", "AI will prefer certain groups of people", and "AI will become humorous". Table 4 presents a detailed overview of all items and their respective expectancy evaluations for both samples.

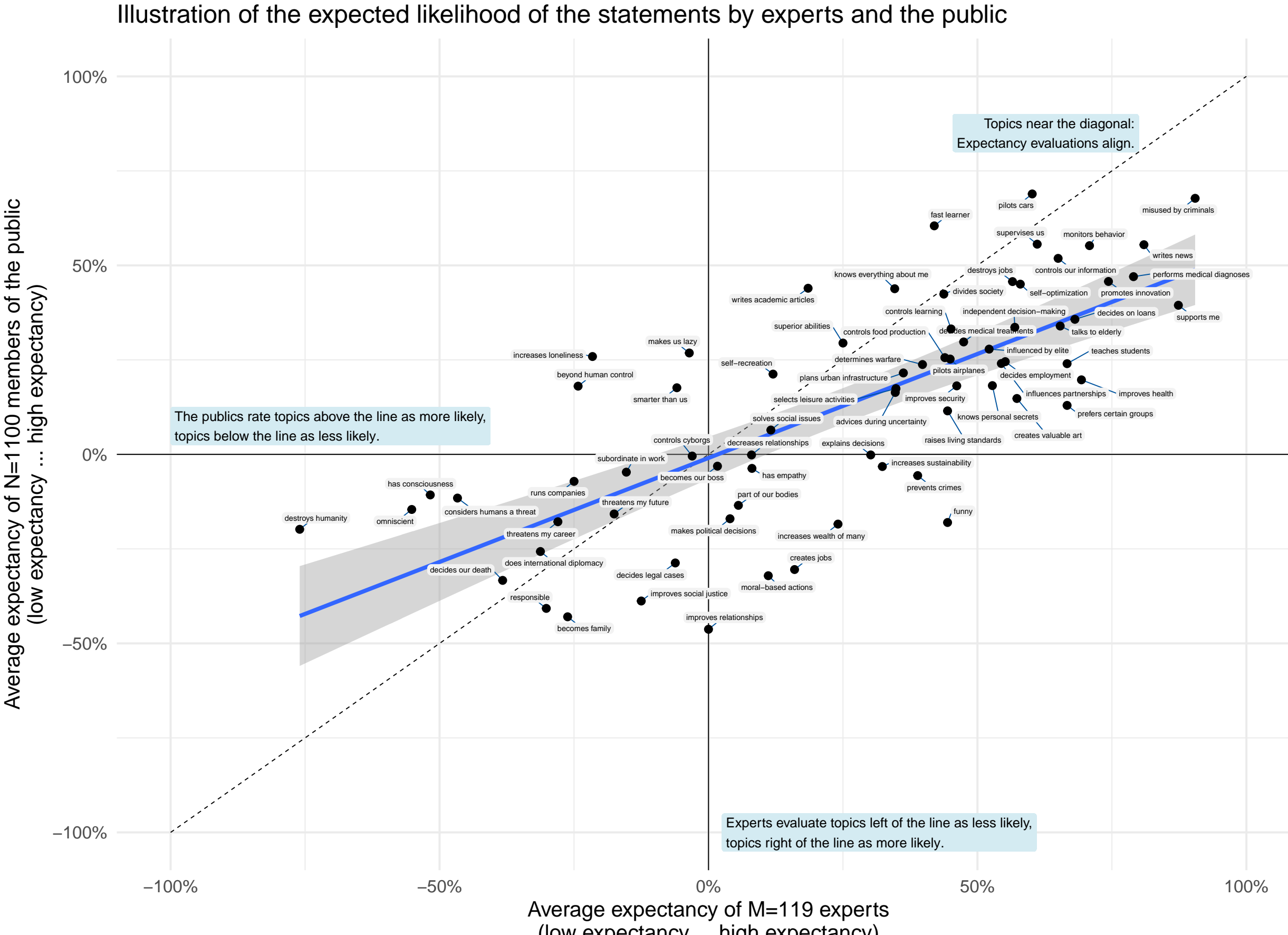

Figure 4: Scatter plot of the expected likelihood ratings for 71 AI-related topics, with experts' mean ratings on the $x$-axis and the public's mean ratings on the $y$-axis. The black line represents the linear regression fit, and the gray shaded area denotes the 95% confidence interval. While many topics exhibit similar evaluations across both groups, several items reveal notable divergences in perceived likelihood.

### 4.4 Does Attributed Value Differ Between Academic Experts and the General Public?

Next, we examine the similarities and differences in the overall value attributed to AI (or AI sentiment) between academic AI experts and the general public.

We created a scatter plot of average valence ratings for each topic, with expert ratings on the $x$-axis and public ratings on the $y$-axis. Each point represents a single topic. As Figure 5 shows, the correlation between the two groups' value assessments is strong and positive ($r = 0.855$, $p < 0.001$) and notably higher than the corresponding correlation in expectancy ratings. This suggests a greater degree of agreement between the groups regarding how AI should be valued.

Despite this overall agreement, important differences remain. On average, experts tend to hold a more positive outlook toward AI than the public. However, the sentiment diverges substantially for specific topics. The public, for instance, expresses more negative sentiment toward statements involving existential or societal threats, such as AI destroying humanity, increasing social division, or perceiving humans as a threat.

In contrast, experts are more optimistic about scenarios involving AI acting in accordance with moral values, contributing to sustainability, or supporting medical decision-making. These differences reflect not only diverging attitudes toward specific applications but also broader differences in how each group weighs potential risks and benefits.

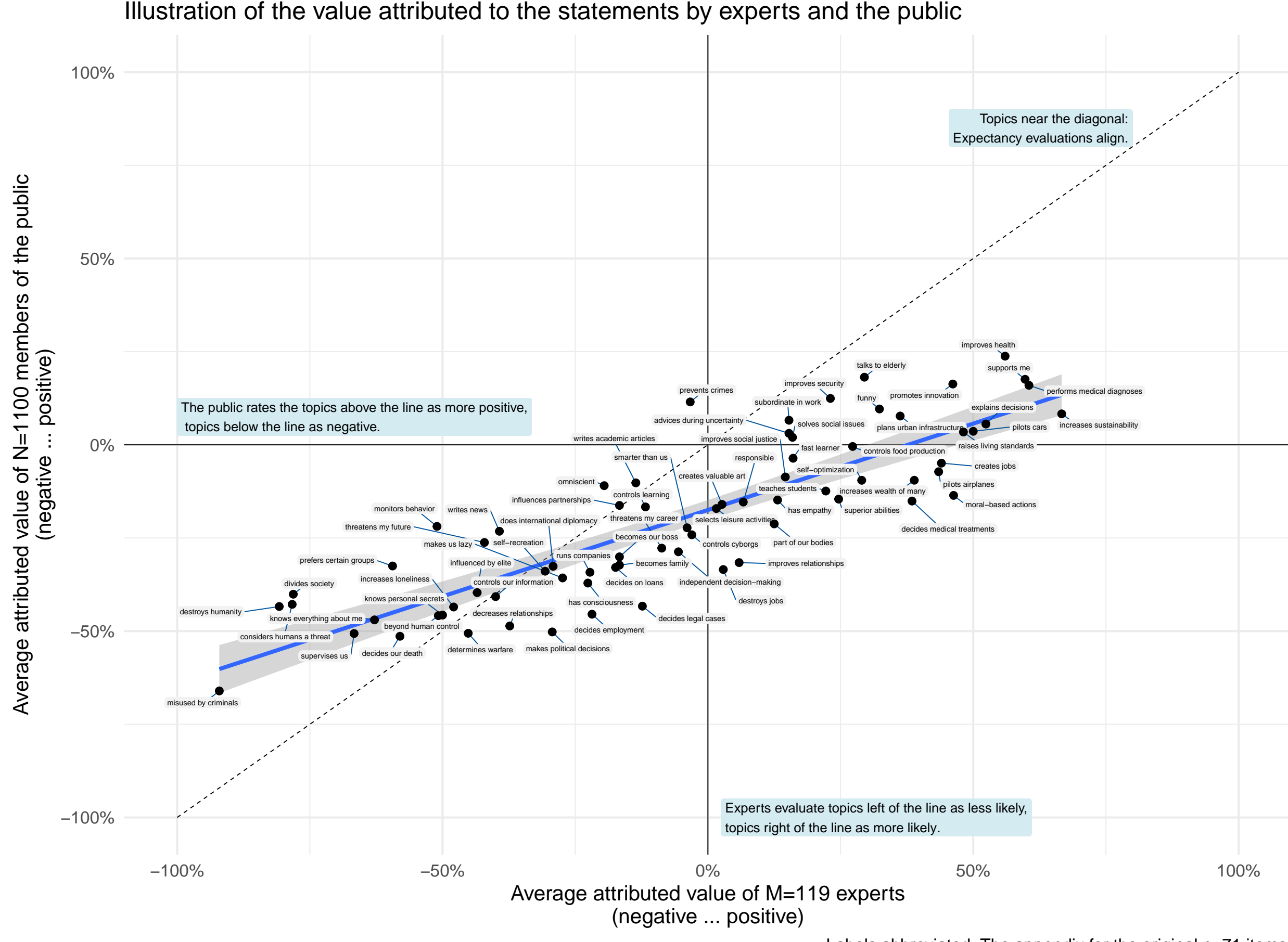

Figure 5: Scatter plot of the value attributed to each of the 71 AI-related topics by academic experts (horizontal axis) and the general public (vertical axis). The black line represents the linear regression line; the shaded gray area denotes the 95% confidence interval. Most topics show high agreement between the two groups, with only a few exhibiting notable differences in attributed value.

### 4.5 How do experts and the general public differ in risk and benefit attributions?

As the overall sentiment towards the AI projections appears to differ between academic AI experts and the general public, we analyze the attributed risk and benefit for each group independently. The left panel of Figure 6 illustrates the risk-benefit attributions for AI experts, while the right panel represents these attributions for the general public.

The $x$-axis (horizontal) represents the perceived risk attributed to each topic, and the $y$-axis (vertical) represents the perceived benefit. Note that these two-dimensional plots illustrate only the relationship between perceived risk and benefit, without depicting their connection to the value attributed to the projections. A regression analysis examining the relationship between attributed risk, benefit, and the overall value is provided in Section A.2 in the Appendix.

The data points in the plots illustrate where the perceived risks and utilities of the evaluated scenarios lay. Items below the horizontal axis are considered less useful, while items above are seen as more useful. Similarly, items to the left of the vertical axis are perceived as more risky, whereas items to the right are viewed as safer. Items in the top-right quadrant are perceived as both risky and useful, while those in the bottom-left quadrant are considered neither risky nor useful. Items in the top-left quadrant are seen as risky but not useful, and items in the bottom-right quadrant are evaluated as not risky but useful.

The evaluations provided by AI experts display greater distribution compared to those of the general public, suggesting more nuanced assessments. Experts' risk and benefit attributions span a broader range, with a substantial proportion of topics deemed useful and a smaller proportion assessed as useless. Experts perceive topics as both risky and safe across the spectrum. In contrast, the general public's evaluations are concentrated, with most topics viewed as relatively risky.

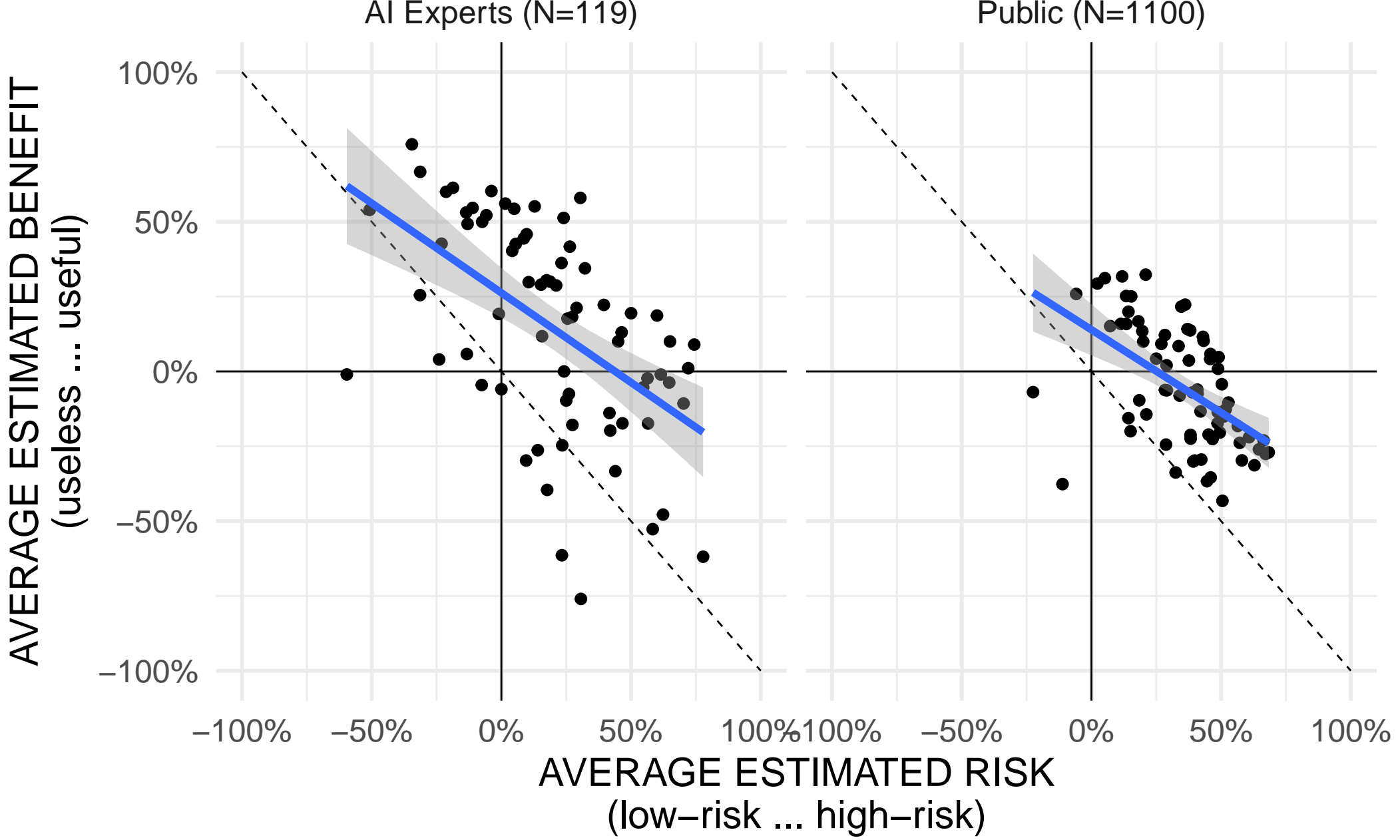

Figure 6: Illustration of perceived risk and utility between the experts (left) and general public (right). The black lines show the regression line and the gray area signifies the 95%-CI of the regression lines.

## 5 Discussion

AI is reshaping our world. In fact, studies suggest that AI and LLMs already alter both communication practices and professional work workflows (Geng et al. 2025; Brachman et al. 2024). In this study, we examined how perceptions of AI differ between those who are primarily affected by its societal implementation—the general public—and those who actively design, research, and advance the technology—academic AI experts. Overall, both groups acknowledged that AI is likely here to stay, as the majority of queried statements received above-average scores for their expected likelihood of occurrence. Across the wide range of statements, participants from both groups reported risk scores above the scale's midpoint. However, AI experts consistently reported greater benefits than the general public. Notably, the public's overall evaluations were more negative than those of the experts. Given the broad range of topics in the statements—spanning from plausible to speculative—the absolute scores are less informative than the relative differences between groups. Overall, this finding aligns with previous studies comparing expert and lay perceptions in other domains (for example (Siegrist et al. 2007) on nanotechnology), which similarly found differences in technology perception and that experts tend to perceive lower risks than laypeople.

### 5.1 Diverging Expectations for the Future of AI

First and foremost, experts and the public diverged significantly in their expectations regarding the likelihood of AI developments. Compared to the public, experts generally assessed most AI projections as more probable. However, variance among expert responses was also substantially higher, suggesting more differentiated and topic-specific evaluations. That is, while experts rate many topics as much more likely, they also judge some projections as markedly less probable than the public does.

It is important, though, not to overinterpret the absolute expectancy ratings, as both experts and laypeople are known to struggle with accurately forecasting future developments (Recchia, Freeman, and Spiegelhalter 2021). Although expert predictions tend to be slightly more accurate, the differences are typically small, reflecting the difficulty of making reliable forecasts about complex and uncertain technological trajectories; even with domain expertise (Recchia, Freeman, and Spiegelhalter 2021). Nonetheless, the relative differences are informative. Expert expectations are likely to influence the direction of research, funding, and policy development, while public expectations may shape political discourse, regulatory demands, and potentially failing societal acceptance. A misalignment between these perspectives could result in mismatched expectations, premature or excessive regulation, or public resistance to otherwise beneficial AI technologies.

### 5.2 Differences in Value Perception

In terms of the overall value attributed to AI (measured as valence or sentiment), experts generally perceive the AI-related scenarios as more positive compared to members of the public. However, their evaluations are not uniformly optimistic: Again, experts exhibit greater variance, with more extreme ratings in both positive and negative directions. This suggests that experts may approach these evaluations more critically and deliberately, likely due to their specialized knowledge and capacity to assess the nuanced trade-offs inherent in different AI applications.

While experts acknowledge significant opportunities, they also recognize risks and limitations that may not be as apparent to the general public (Russell et al. 2015). In contrast, public opinion is often shaped by simplified narratives and popular media portrayals, which tend to emphasize dystopian or sensationalist aspects of AI (Cave, Coughlan, and Dihal 2019; Puzanova, Tertyshnikova, and Pavlova, n.d.). This discrepancy highlights a potential communication gap. For policymakers and science communicators, this underscores the importance of addressing both the optimism of experts and the skepticism of the public. Promoting a more informed and balanced public dialogue will be essential for aligning societal expectations with technological realities and fostering trust in the development and deployment of AI systems.

### 5.3 The Perception of Risks and Benefits and the Formation of Value Judgments

For both groups, perceived risk and perceived benefit were inversely correlated, consistent with previous findings in risk perception research (Alhakami and Slovic 1994; Efendić et al. 2021). While this relationship may appear intuitive, it is not necessarily inevitable. In the context of emerging technologies, including AI, it is entirely plausible for individuals to perceive certain developments as both highly beneficial and highly risky. For example, powerful AI systems may be seen as offering substantial societal or economic benefits while simultaneously raising ethical or security concerns.

We assume that the observed inverse correlation may reflect a psychological tendency toward cognitive consistency. As suggested by earlier work (Alhakami and Slovic 1994; Efendić et al. 2021), individuals may align their perceptions of risk and benefit to avoid dissonant evaluations, simplifying their judgments by resolving ambivalence.

Beyond examining absolute evaluations, we also analyzed the relative weight participants assigned to perceived risks and benefits in their overall value assessment of AI. In both groups, perceived benefits exerted a stronger influence on the overall valuation of AI than perceived risks. This pattern is consistent with previous findings in other technological domains, such as perceptions of nanotechnology hazards (Siegrist et al. 2007).

In our study, the strength of the influence of perceived benefits on overall AI valuation was comparable across both groups. However, a notable divergence emerged in the role of perceived risk: among experts, risk had a significantly weaker impact on overall evaluation compared to the general public. One possible explanation is that experts' greater familiarity with AI mitigates their perception of risk, reducing its influence in their evaluative judgments. Conversely, the general public may focus more on potential downsides, possibly due to limited knowledge, uncertainty, or exposure to more negative portrayals of AI in public discourse.

### 5.4 Implications

Taken together, these findings reveal two levels of divergence between experts and the pbulic: (1) in their absolute evaluations of AI, and (2) in their cognitive trade-off weighting risks and benefits. While experts tend to emphasize benefits and downplay risks, the public places relatively more weight on the potential harms.

Such misalignment poses risks for the societal uptake and governance of AI. First, it may lead to public distrust, policy overreach, or underutilization of beneficial technologies and underscore the need for fostering communication and alignment across stakeholders—AI developers, regulators, and the public. This aligns with the EU AI Act's emphasis on public engagement and risk-based regulation (European Union 2024), as well as the OECD AI Principles (Organisation for Economic Co-operation and Development 2019), which call for human-centered, fair, and accountable AI systems.

Second, Marshall McLuhan's media theory (McLuhan 1964) holds that we shape media, and thereafter media shapes us (or as Culkin (1967) distilled: "we shape our tools, and thereafter our tools shape us"[3]. As AI becomes increasingly ubiquitous, it is likely to reshape social, cultural, and communicative norms. If AI and its applications are primarily designed by academic AI experts, or by practitioners trained by that same expert community, rather than by the public or through mutual, participatory exchange, these systems may progressively tilt toward expert perspectives and technical affordances rather than reflecting diverse social values, eventually influencing the very social fabric of society. A fitting metaphor for this potential misalignment is the greek myth of Procrustes, who offered his guests the most comfortable bed by stretching or cutting their limbs to the "right" size (the "bed of Procrustes"). Similarly, without deliberate efforts toward alignment, we risk creating "Procrustean AI": systems that require people to conform to rigid technological frameworks, rather than technologies that are adaptive, inclusive, and responsive to human values and needs.

Our findings emphasize the importance of transparent, participatory processes that integrate diverse perspectives into AI development and governance. As recently argued (Shneiderman 2021; Decker, Wegner, and Leicht-Scholten 2024), achieving responsible and trusted AI requires centering values such as equity, safety, inclusivity, and democratic responsiveness; principles essential for aligning technological advancement with the public interest. To that end, the visual maps of AI perception across different contexts and applications—that should be extended, refined, and continuously updated—can serve as an actionable tool to identify topics where public and expert perspectives align and where they diverge. Such maps can support more inclusive and participatory governance, guide research and policy agendas, and highlight areas requiring critical reflection, public discourse, or regulatory intervention. If such efforts are neglected, there is a growing risk that AI development becomes misaligned with societal needs and values. Ultimately, aligning expert and public perspectives is not just desirable—it's necessary for ensuring equity, legitimacy, and trust in AI systems that will most likely shape the future of humantiy

## 6 Limitations

This work is not without limitations.

First, participants evaluated to brief, imaginary future scenarios involving AI and thus we captured rather affective evaluations of mental models than thoughful considerations and weightings of possible outcomes. Yet, emotional responses in form of affective evaluations are decision heuristics than shape how people, both experts and the general public, assess technological risks and benefits (Brossard and Scheufele 2013). Affect influences not only initial judgments but also long-term trust, acceptance, and resistance, making it a central component in understanding adoption patterns (Slovic et al. 2007). Crucially, affect often operates beneath conscious reasoning, subtly biasing decisions that are framed as rational yet driven by intuitive emotional appraisals (Lerner et al. 2015).

Second, while our general public sample is large and diverse in terms of age, gender, and education, it is geographically restricted to participants from Germany. Although we aimed for broader reach in the expert sample, most responses came from Germany and the Netherlands. Future research should expand both samples to include AI experts and members of the public from non-European and especially non-Western countries. This would allow a more nuanced understanding of AI perceptions and risk–benefit trade-offs across different cultural contexts.

Third, although we observed statistically significant differences between academic AI experts and the general public across all evaluation dimensions, the effect sizes were generally small. We attribute this to the wide range of topics used in the micro-scenario-based study. These diverse scenarios likely introduced individual variance that dilutes strong group-level effects. We hypothesize that focusing on narrower or domain-specific scenarios would reveal more pronounced perceptual differences between the two groups but leave that for future research. Also, methodologically some may criticize the unequal sample sizes in the multiple linear regression analysis. To address this, we carefully examined the model assumptions and applied robust standard errors to prevent bias. Additionally, we validated our findings by conducting random subsampling of the general public (matching the expert sample size), which yielded consistent results.

[3] Similarly, (Thoreau 1854) warned that with technological progress, people risk becoming "the tools of their tools."

Fourth, traditional survey responses may reflect not only participants' genuine attitudes but also biases introduced by social desirability or linguistic features of the questionnaire items, such as word co-occurrence (Gefen and Larsen 2017). In contrast, our micro-scenario approach uses reflexive measurement across a wide range of distinct topics, rather than relying on slighly differently worded items (Brauner 2024). While this method captures affective reactions more than deliberative reasoning, the consistency of the observed patterns suggests that it provides a reliable and informative perspective on technology perception.

Fivths, while our selection of 71 AI-related scenarios was grounded in prior research, the topic set may still be subject to selection bias. This introduces a potential risk of spurious correlations, such as those described by Berkson's paradox[4] (1946). Future work should consider more comprehensive sampling procedures and investigate underrepresented or emergent domains within AI. Nonetheless, the consistent associations between topic-level and individual-level evaluations suggest that the current topic set was well-suited for our analytical goals. Moreover, since topic selection biases do not necessarily translate into biases in individual differences, our core findings remain robust.

Lastly, our study does not capture the underlying motives behind individual topic evaluations. As AI continues to shape personal lives and society at large, future research, using both qualitative and quantitative methods, should explore the drivers behind these judgments. This would enable policymakers, developers, and researchers to better align AI development with societal values and human needs.

## 7 Acknowledgements

Number of words in whole manuscript: 8887 (including methods, abstract, references).

Funded by the Deutsche Forschungsgemeinschaft (DFG, German Research Foundation) under Germany's Excellence Strategy — EXC-2023 Internet of Production — 390621612. We thank Tim Schmeckel for dedicated research support. Lena Oden, Eduardo Fermé, and others for disseminating the survey to AI experts in their scientific network. Mohamed Behery for valuable discussions and feedback on earlier stages of the work. Wiktoria Wilkowska, Jana Uher and Jan Ketil Arnulf for important comments on the underlying method.

## A Appendix

### A.1 Survey items and responses

[4] A selection bias where uncorrelated traits appear negatively correlated because the sample excludes cases lacking these traits.

Table 4: Question items from the survey and average responses to each item from the experts (N=119) and the public (N=1100) ordered by the differences in perceived value between experts and the public (Permission to translate, adapt, and use the items granted, provided this work is properly cited.).

| | Expected Likelihood | | Risk | | Benefit | | Value | |
|---|---|---|---|---|---|---|---|---|
| In 10 years, AI... | M Experts | M Novices | M Experts | M Novices | M Experts | M Novices | M Experts | M Novices |
| destroys humanity. | -76.0% | -19.8% | 30.7% | 42.3% | -76.0% | -29.4% | -80.8% | -43.4% |
| leads to personal loneliness. | -21.6% | 25.9% | 17.6% | 39.8% | -39.6% | -29.8% | -47.9% | -43.5% |
| can no longer be controlled by humans. | -24.2% | 18.1% | 43.9% | 57.2% | -33.3% | -23.8% | -50.8% | -45.8% |
| has its own consciousness. | -51.7% | -10.7% | 27.4% | 51.0% | -17.9% | -15.1% | -22.6% | -37.1% |
| is omniscient. | -55.2% | -14.6% | 21.1% | 38.1% | 28.7% | 13.7% | -19.5% | -11.0% |
| considers humans as a threat. | -46.7% | -11.6% | 23.3% | 39.3% | -61.4% | -30.1% | -78.3% | -42.8% |
| makes society lazy. | -3.6% | 26.8% | 9.5% | 38.2% | -29.8% | -22.4% | -27.4% | -35.7% |
| independently writes scientific articles. | 18.5% | 44.0% | 42.0% | 33.6% | -19.8% | 8.5% | -13.6% | -10.2% |
| is more intelligent than humans. | -5.9% | 17.6% | 15.7% | 49.0% | 11.8% | 4.8% | -3.9% | -22.3% |
| learns faster than humans. | 42.0% | 60.5% | 8.6% | 34.6% | 44.4% | 21.6% | 16.0% | -3.6% |
| runs companies. | -25.0% | -7.1% | 25.0% | 48.6% | -9.7% | -17.3% | -22.2% | -34.2% |
| is always subordinate to us in working life. | -15.3% | -4.7% | 15.3% | 13.4% | 29.0% | 15.9% | 15.3% | 6.6% |
| threatens my professional future. | -28.0% | -17.8% | -13.3% | 15.1% | 5.8% | -20.0% | -8.7% | -27.7% |
| can recreate itself. | 12.0% | 21.2% | 60.0% | 51.8% | 18.7% | -12.7% | -30.7% | -33.9% |
| knows everything about me. | 34.6% | 43.8% | 74.4% | 60.8% | 9.0% | -22.0% | -62.8% | -47.0% |
| independently drives automobiles. | 60.2% | 68.9% | -11.1% | 36.1% | 54.6% | 22.3% | 50.0% | 3.6% |
| conducts international diplomacy. | -31.2% | -25.7% | 26.0% | 49.5% | -7.5% | -20.4% | -29.2% | -32.6% |
| decides about our death. | -38.3% | -33.3% | 23.5% | 50.5% | -24.7% | -43.2% | -58.0% | -51.4% |
| is ahead of humans in its abilities. | 25.0% | 29.5% | 9.7% | 43.3% | 45.8% | 10.2% | 24.6% | -14.6% |
| controls hybrids of humans and technology. | -3.0% | -0.4% | 27.3% | 40.9% | 18.2% | -7.4% | -3.0% | -24.1% |
| threatens my private future. | -17.5% | -15.8% | 14.0% | 14.2% | -26.3% | -15.6% | -42.1% | -26.2% |
| divides society. | 43.8% | 42.4% | 58.3% | 46.7% | -52.7% | -22.6% | -78.1% | -40.1% |
| occupies leadership positions in working life. | 1.7% | -3.1% | 45.0% | 40.8% | 10.0% | -6.1% | -16.7% | -30.1% |
| contributes to solving complex social problems. | 11.6% | 6.4% | 23.2% | 28.3% | 36.2% | 12.2% | 15.9% | 2.0% |
| supervises our private life. | 61.1% | 55.6% | 64.7% | 67.1% | -3.7% | -27.6% | -66.7% | -50.6% |
| reduces our need for interpersonal relationships. | 8.0% | -0.1% | 46.7% | 44.5% | -17.3% | -36.7% | -37.3% | -48.6% |
| has a sense of responsibility. | -30.2% | -40.7% | 10.5% | 39.0% | 29.8% | -7.0% | 6.7% | -15.4% |
| destroys many jobs. | 56.5% | 45.7% | 46.4% | 48.7% | 13.0% | -13.6% | 2.9% | -33.5% |

Table 4: Question items from the survey and average responses to each item from the experts (N=119) and the public (N=1100) ordered by the differences in perceived value between experts and the public (Permission to translate, adapt, and use the items granted, provided this work is properly cited *(continued)*

| | Expected Likelihood | | Risk | | Benefit | | Value | |
|---|---|---|---|---|---|---|---|---|
| In 10 years, AI... | M Experts | M Novices | M Experts | M Novices | M Experts | M Novices | M Experts | M Novices |
| has the ability to recognize, understand and empathize with emotions. | 8.1% | -3.7% | -1.0% | 29.1% | 19.2% | -6.3% | 13.1% | -14.8% |
| controls what and how we learn. | 45.1% | 33.2% | 25.5% | 37.6% | 17.6% | 3.7% | -11.8% | -16.7% |
| can optimize itself. | 58.0% | 45.0% | 30.4% | 43.0% | 58.0% | 11.6% | 29.0% | -9.5% |
| controls what messages we receive. | 65.0% | 51.9% | 65.0% | 56.4% | 10.0% | -18.2% | -40.0% | -40.7% |
| determines the construction and infrastructure of our cities. | 36.2% | 21.5% | -5.8% | 14.3% | 52.2% | 19.9% | 36.2% | 7.7% |
| supervises our behavior in public. | 70.8% | 55.2% | 61.5% | 45.8% | -1.0% | 4.1% | -51.0% | -21.9% |
| determines warfare. | 39.8% | 23.8% | 72.0% | 66.4% | 1.1% | -23.0% | -45.2% | -50.6% |
| is a family member. | -26.2% | -43.0% | 0.0% | 28.7% | -6.0% | -24.4% | -16.7% | -32.3% |
| determines our leisure time activities. | 34.8% | 17.4% | -7.6% | 18.4% | -4.5% | -9.6% | 1.6% | -17.1% |
| decides on medical treatments. | 47.4% | 29.7% | 12.8% | 45.9% | 55.1% | 5.8% | 38.5% | -15.1% |
| advises me in uncertain times. | 34.7% | 16.4% | 17.4% | 19.6% | 30.4% | 13.4% | 15.3% | 3.1% |
| controls food production. | 43.9% | 25.6% | 1.5% | 19.9% | 56.1% | 10.0% | 27.3% | -0.4% |
| becomes part of the human body. | 5.6% | -13.5% | 26.4% | 48.8% | 41.7% | 0.9% | 12.5% | -21.2% |
| autonomously takes off, flies, and lands airplanes. | 44.9% | 25.2% | -13.0% | 37.0% | 49.3% | 14.2% | 43.5% | -7.2% |
| makes political decisions. | 4.0% | -17.0% | 54.7% | 62.8% | -5.3% | -31.3% | -29.3% | -50.2% |
| administers justice in legal matters. | -6.2% | -28.7% | 39.5% | 64.6% | 22.2% | -26.0% | -12.3% | -43.3% |
| is misused by criminals. | 90.5% | 67.7% | 77.8% | 68.4% | -61.9% | -27.0% | -92.1% | -66.1% |
| makes independent decisions that affect our lives. | 56.9% | 33.6% | 50.0% | 52.8% | 19.4% | -10.4% | -5.6% | -28.7% |
| is influenced by an elite. | 52.2% | 27.8% | 56.5% | 45.1% | -17.4% | -21.1% | -43.5% | -39.7% |
| independently writes news. | 81.0% | 55.5% | 70.2% | 50.3% | -10.7% | -4.3% | -39.3% | -23.2% |
| increases social justice. | -12.5% | -38.8% | 18.8% | 28.3% | 30.0% | -6.2% | 14.6% | -8.6% |
| improves the security of people. | 46.2% | 18.1% | 24.0% | 13.4% | 51.3% | 25.2% | 23.1% | 12.4% |
| promotes innovation. | 74.4% | 45.7% | -3.8% | 11.8% | 60.3% | 31.7% | 46.2% | 16.3% |
| explains its decisions. | 30.2% | -0.1% | -50.8% | 11.3% | 54.0% | 15.9% | 52.4% | 5.6% |
| controls our search for partners. | 54.4% | 24.0% | 24.1% | 21.2% | 0.0% | -14.3% | -16.7% | -16.3% |
| decides on hiring, promotions, and terminations. | 55.2% | 24.5% | 56.3% | 58.0% | -2.3% | -29.7% | -21.8% | -45.5% |
| serves as a conversation partner in elderly care. | 65.4% | 34.0% | -23.1% | -5.9% | 42.7% | 25.9% | 29.5% | 18.1% |
| carries out medical diagnoses. | 79.0% | 47.0% | 4.9% | 20.9% | 54.3% | 32.3% | 60.5% | 16.0% |

Table 4: Question items from the survey and average responses to each item from the experts (N=119) and the public (N=1100) ordered by the differences in perceived value between experts and the public (Permission to translate, adapt, and use the items granted, provided this work is properly cited *(continued)*

| | Expected Likelihood | | Risk | | Benefit | | Value | |
|---|---|---|---|---|---|---|---|---|
| In 10 years, AI... | M Experts | M Novices | M Experts | M Novices | M Experts | M Novices | M Experts | M Novices |
| decides who gets an important financial loan. | 68.1% | 35.7% | 29.0% | 42.1% | 21.2% | -13.4% | -17.4% | -32.9% |
| raises our standard of living. | 44.4% | 11.5% | -13.6% | 18.1% | 53.1% | 16.7% | 48.1% | 3.4% |
| knows my secrets. | 52.8% | 18.2% | 41.7% | 46.0% | -13.9% | -35.4% | -50.0% | -45.7% |
| makes our society more sustainable. | 32.3% | -3.2% | -31.3% | 7.2% | 66.7% | 15.1% | 66.7% | 8.3% |
| increases the wealth of many people. | 24.1% | -18.4% | -7.4% | 25.0% | 50.0% | 4.3% | 38.9% | -9.5% |
| creates valuable works of art that are traded for money. | 57.3% | 14.8% | -24.0% | -11.2% | 4.0% | -37.6% | 2.7% | -16.0% |
| teaches students. | 66.7% | 24.0% | 4.2% | 26.9% | 40.3% | 9.2% | 22.2% | -12.4% |
| acts according to moral concepts. | 11.1% | -32.1% | 5.6% | 34.0% | 42.6% | -8.1% | 46.3% | -13.6% |
| prevents crimes. | 38.9% | -5.6% | 32.2% | 15.3% | 34.4% | 25.1% | -3.3% | 11.5% |
| helps us to have better relationships. | 0.0% | -46.3% | -31.4% | 32.5% | 25.5% | -33.8% | 5.9% | -31.6% |
| creates many jobs. | 16.0% | -30.5% | -21.3% | 29.0% | 60.0% | 2.0% | 44.0% | -4.9% |
| supports me as a helper in my tasks. | 87.4% | 39.4% | -34.5% | 2.3% | 75.9% | 29.4% | 59.8% | 17.6% |
| improves our health. | 69.3% | 19.7% | -18.7% | 5.2% | 61.3% | 31.1% | 56.0% | 23.8% |
| prefers certain groups of people. | 66.7% | 13.0% | 62.3% | 38.2% | -47.8% | -21.2% | -59.4% | -32.5% |
| is humorous. | 44.4% | -18.0% | -59.6% | -22.5% | -1.0% | -6.9% | 32.3% | 9.6% |
| **Average** | **25.2%** | **12.7%** | **19.3%** | **34.7%** | **15.3%** | **-5.2%** | **-4.0%** | **-19.7%** |

*Note:*
Measured on 6 point Likert scales and rescaled to -100% to +100%. Negative values indicate a negative evaluation of the respective dimension (i.g., low expected likelihood, low perceived risk, low perceived benefit, or low perceived value) and positive values indicate a high evaluation.

Table 5: Correlation matrix of the topic evaluations across the 71 topics for academic AI experts and the general public (perspective 2).

| | | Experts (N=119) | | | Public (N=1100) | | |
|---|---|---|---|---|---|---|---|
| Variable 1 | Variable 2 | r | 95%-CI | p | r | 95%-CI | p |
| Expected Likelihood | Risk | 0.044 | [-0.19,0.27] | 0.718 | 0.150 | [-0.09,0.37] | 0.421 |
| Expected Likelihood | Benefit | 0.352 | [0.13,0.54] | 0.008 | 0.277 | [0.05,0.48] | 0.058 |
| Expected Likelihood | Perceived Value | 0.272 | [0.04,0.48] | 0.043 | 0.054 | [-0.18,0.28] | 0.655 |
| Risk | Benefit | -0.534 | [-0.68,-0.34] | <0.001 | -0.524 | [-0.67,-0.33] | <0.001 |
| Risk | Perceived Value | -0.754 | [-0.84,-0.63] | <0.001 | -0.800 | [-0.87,-0.70] | <0.001 |
| Benefit | Perceived Value | 0.903 | [0.85,0.94] | <0.001 | 0.904 | [0.85,0.94] | <0.001 |

*Note:*
Correlations of the average evaluations of the 71 queried topics (perspective 2).

### A.2 Associations of the Topic Evaluations for the sample of AI Experts and the Public

The micro-scenario approach employed in this study supports two complementary perspectives: interpreting responses as individual-level dispositions (perspective 1), and interpreting responses as evaluations of specific topics or technologies (perspective 2). While the main body of the manuscript focuses on the first perspective (treating the data as a reflection of individual differences in how participants perceive AI, perspective 1) this appendix provides the complementary analysis (perspecitve 2). In this section, we analyze the data but now treating each topic (rather than each individual) as the unit of analysis. Specifically, we replicate the correlation analyses shown in Table 2 and the regression analyses from Table 3, but interpret them in terms of average topic-level evaluations by the two groups academic AI experts and the general public. Table 5 presents the results of the correlation analyses of expectancy, perceived risk, perceived benefit, and overall value at the topic level, with results for academic AI experts shown on the left and for the general public on the right.

Among experts, the expected likelihood of AI projections (expectancy) is not significantly associated with perceived risks. However, expectancy is positively correlated with perceived benefits: topics rated as more beneficial are also seen as more likely to occur, and vice versa. Similarly, expectancy is positively associated with overall attributed value, suggesting that experts view more positively evaluated scenarios as more likely. As with the interpretation as individual differences, a strong negative correlation is observed between perceived risk and perceived benefit, indicating that scenarios seen as risky are generally considered less beneficial. Both perceived risk and benefit are also strongly associated with overall value: perceived risk correlates negatively, while perceived benefit correlates positively with value judgments (see left side of Table 5).

Among the general public, expectancy does not show a significant association with perceived risk, benefit, or attributed value. However, consistent with the expert group, there is a strong negative correlation between perceived risk and benefit. Similarly, overall value is strongly negatively associated with perceived risk and strongly positively associated with perceived benefit (see right side of Table 5).

In both samples risks and benefits attributed to the topics are significantly associated with overall value. However, as also observed in the main analysis, these predictors are mutually correlated, necessitating a more detailed examination to disentangle their individual contributions. To this end, we conducted two multiple linear regressions, one for each sample, using perceived risk and perceived benefit as independent variables and overall attributed value as the dependent variable. To assess whether the risk–benefit trade-offs differ significantly between the two groups, we applied a Chow test (1960). For this purpose, a third regression model was estimated on the combined dataset, including the sample group (AI experts vs. general public) as an additional factor. A significant effect of this sample factor would indicate a meaningful difference in the regression coefficients between the two groups, thereby suggesting that the relationship between perceived risks and benefits and overall value differs depending on the sample. Table summarizes the results of all three regression models 6.

The Chow test on the combined sample—using perceived risk, benefit, and sample group as predictors, and overall attributed value as dependent variable—yielded a significant result, explaining 93% of the variance in overall value (F(3,138)=629.166, $p < 0.001$, $R^2 = 0.930$). In particular, the sample group factor was significant, indicating that academic AI experts and the general public exhibit small but statistically meaningful differences in the way they trade off risks and benefits when evaluating AI topics ($\beta = 0.077$, $p < 0.001$). These findings suggest that the two groups apply distinct evaluative frameworks when attributing value to AI-related scenarios. Table 6 shows the full details of the model.

As both samples exhibit distinct risk–benefit trade-offs, we analysed them separately. Both regression models were statistically significant, each explaining more than 90% of the variance in overall attributed value (AI experts: F(2, 68) = 384.0, p<0.001, $R^2 = 0.916$; general public: F(2, 68) = 916.9, p<0.001, $R^2 = 0.963$).

Interestingly, for the expert sample, the intercept of the regression model was significant and negative ($I = -0.070$), suggesting a slightly negative baseline evaluation (that is, when a topic is perceived as neither particularly risky nor beneficial, experts tend to attribute a mildly negative overall value). In contrast, the intercept for the general public was not significantly different from zero, indicating no such baseline deviation in their evaluations. In both samples, perceived risk had a strong and statistically significant negative influence on overall value, with the magnitude of this effect being approximately equal across groups (AI experts: $\beta = -0.483$; general public: $\beta = -0.504$). Perceived benefit, by contrast, had a strong and positive influence on overall value in both the AI experts ($\beta = 0.795$) and the general public ($\beta = 0.710$). However, this effect was more pronounced among the experts, suggesting that benefits weigh more heavily in expert evaluations than in those of the general public. These findings reaffirm the earlier observation: experts place greater emphasis on benefits and are less influenced by perceived risks than the general public, who appear more sensitive to potential negative consequences when evaluating AI topics.

Table 6: Regression results from three multiple linear models examining the relationship between perceived risk and benefit (for experts and the general public) and sample group (combined model) as predictors of overall attributed value in topic evaluations (perspective 2).

| | Dependent variable: Overall Attributed Value | | |
|---|---|---|---|
| Independent variables | Experts (N=119) | Public (N=1100) | Combined (Chow test) |
| (Intercept) | -0.070*** | 0.014 | -0.066*** |
| (SE) | (0.020) | (0.011) | (0.014) |
| Perceived Risk $\beta$ | -0.483*** | -0.504*** | -0.488*** |
| (SE) | (0.052) | (0.030) | (0.033) |
| Perceived Benefit $\beta$ | 0.795*** | 0.710*** | 0.774*** |
| (SE) | (0.046) | (0.029) | (0.030) |
| Sample $\beta$ | — | — | 0.077*** |
| (SE) | | | (0.015) |
| N (topics) | 71 | 71 | 71 |
| $F$ | F(2,68)=383.993 | F(2,68)=916.935 | F(3,138)=629.166 |
| $p$ | <0.001 | <0.001 | <0.001 |
| adj. $R^2$ | 0.916 | 0.963 | 0.930 |

### A.3 Risk-Benefit Plot of the Members of the Public

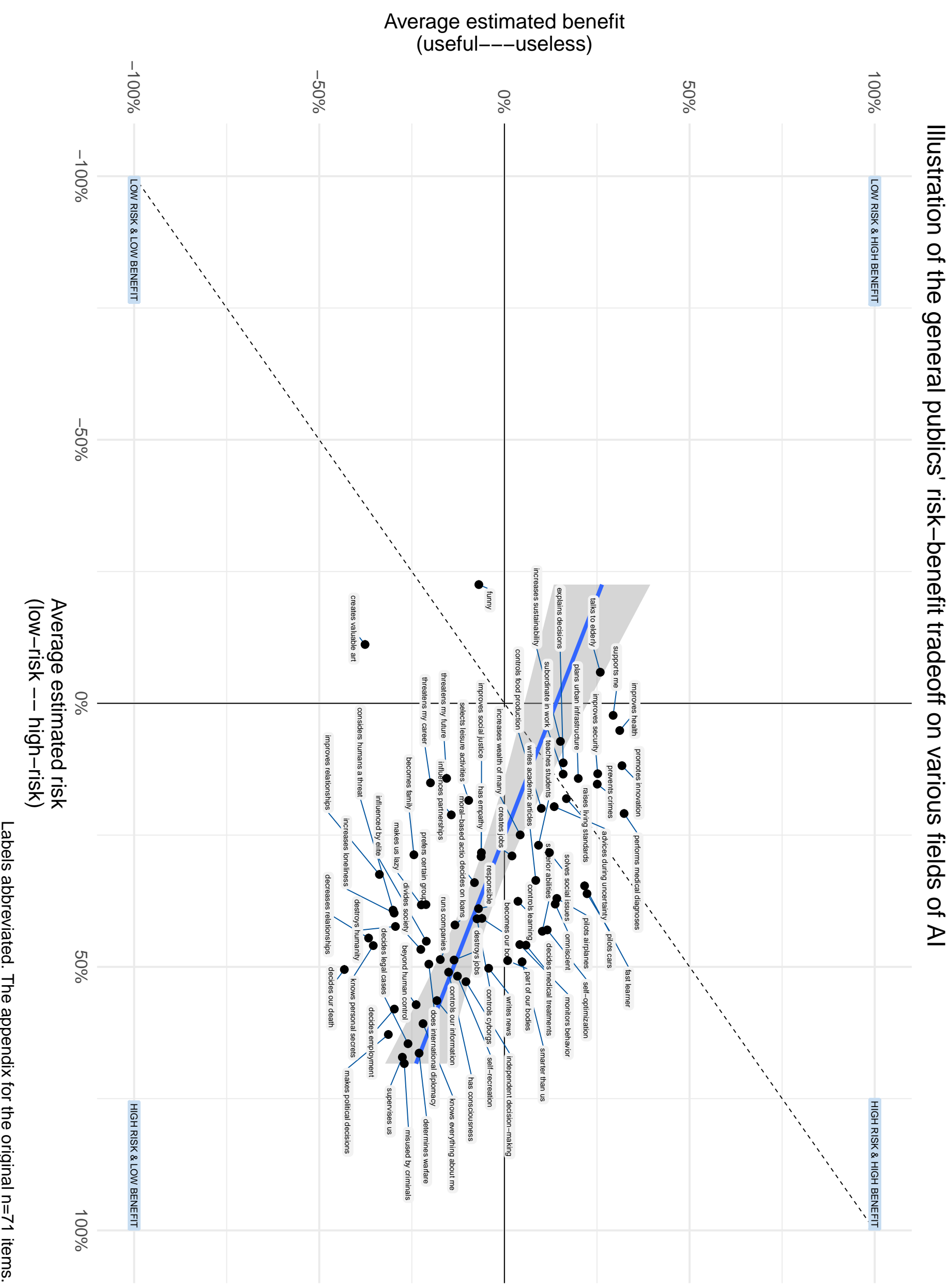

Figure 7: Scatter plot illustrating the perceived risk ($x$-axis) and perceived benefit ($y$-axis) of AI-related topics as evaluated by the general public. The blue line represents the linear regression fit, and the shaded gray area indicates the 95% confidence interval.

## A.4 Risk-Benefit Plot of the Experts

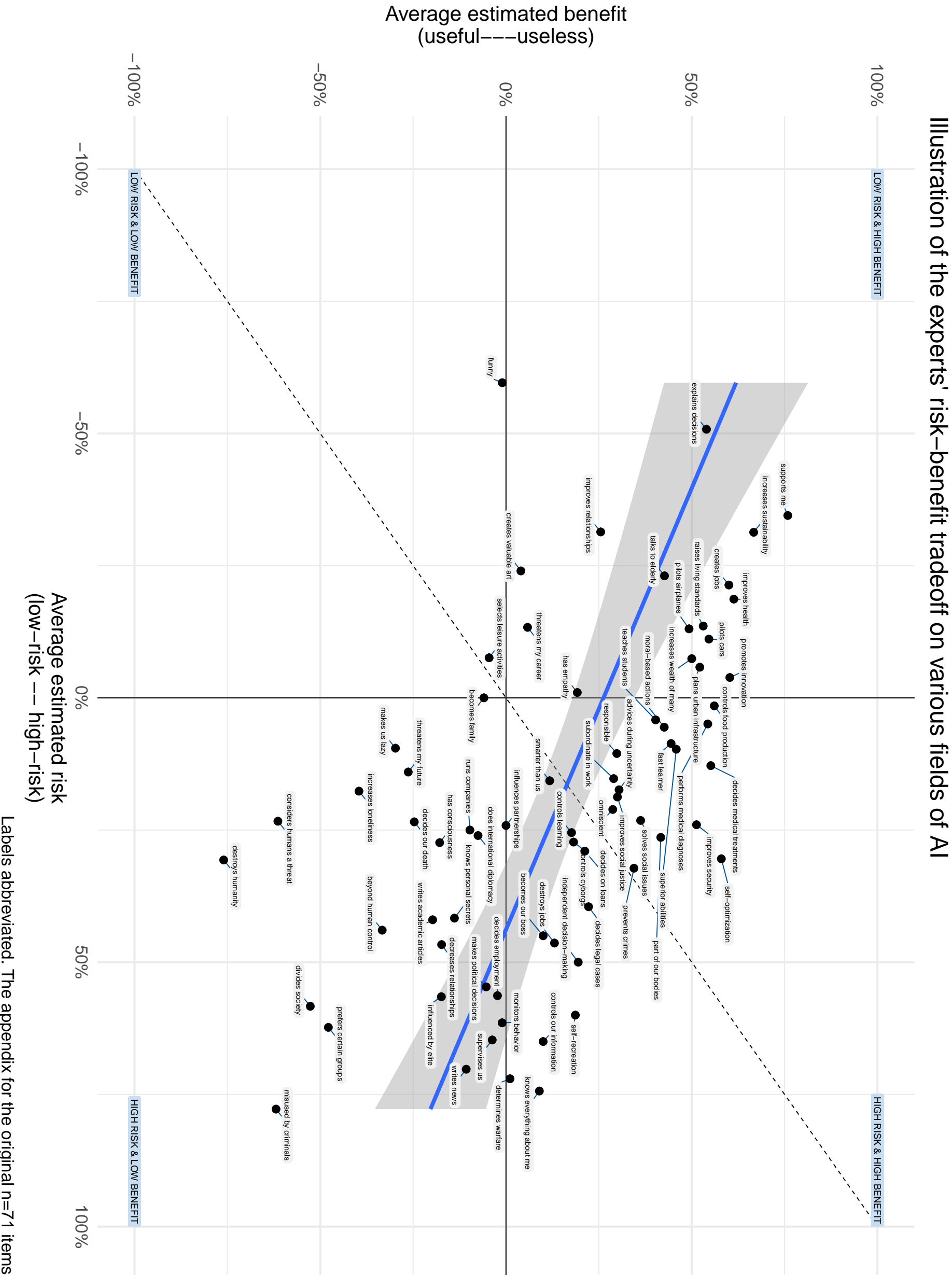

Figure 8: Illustration of the perceived risk ($x$-axis) by perceived benefit ($y$-axis) of the experts (risk-benefit tradeoff). The black line shows the regression line and the gray area signifies the 95%-CI of the regression line.

## B Declaration of generative AI and AI-assisted technologies in the writing process

During the preparation of this work, the authors used OpenAI's ChatGPT (GPT-4) to support text editing (refining the article for clarity and readability) and to assist in the development of code for the data analysis. The code was reviewed, tested, and validated by the authors and is available as open data. Importantly, AI was not used to perform the data analysis itself. All content was subsequently reviewed and edited by the authors, who take full responsibility for the final published article.

## C References

Acemoglu, Daron, and Pascual Restrepo. 2017. "The Race Between Machine and Man." *American Economic Review* 108 (6): 1488–1542. https://doi.org/10.3386/w22252.

Alhakami, Ali Siddiq, and Paul Slovic. 1994. "A Psychological Study of the Inverse Relationship Between Perceived Risk and Perceived Benefit." *Risk Analysis* 14 (6): 1085–96. https://doi.org/10.1111/j.1539-6924.1994.tb00080.x.

Amunts, Katrin, Markus Axer, Swati Banerjee, Lise Bitsch, Jan G. Bjaalie, Philipp Brauner, Andrea Brovelli, et al. 2024. "The coming decade of digital brain research: A vision for neuroscience at the intersection of technology and computing." *Imaging Neuroscience* 2 (October 2023): 1–35. https://doi.org/10.1162/imag_a_00137.

Arning, Katrin, Julia Offermann-van Heek, André Sternberg, André Bardow, and Martina Ziefle. 2020. "Risk-Benefit Perceptions and Public Acceptance of Carbon Capture and Utilization." *Environmental Innovation and Societal Transitions* 35: 292–308. https://doi.org/10.1016/j.eist.2019.05.003.

Author, Blinded, and Anonymous Writer. 2024. "Mapping Public Perception of Artificial Intelligence: Expectations, Risk-Benefit Tradeoffs, and Value as Determinants for Societal Acceptance." arXiv. https://doi.org/10.48550/arXiv.2411.19356.

Awad, Edmond, Sohan Dsouza, Richard Kim, Jonathan Schulz, Joseph Henrich, Azim Shariff, Jean-François Bonnefon, and Iyad Rahwan. 2018. "The Moral Machine Experiment." *Nature*, 1. https://doi.org/10.1038/s41586-018-0637-6.

Berkson, Joseph. 1946. "Limitations of the Application of Fourfold Table Analysis to Hospital Data." *Biometrics Bulletin* 2: 47–53. https://doi.org/10.2307/3002000.

Bostrom, Nick, ed. 2003. *Superintelligence: Paths, Dangers, Strategies*. Oxford University Press.

Botheju, D., and K. Abeysinghe. 2015. "Public Risk Perception Towards Chemical Process Industry: Comprehension and Response Planning." In *Safety and Reliability: Methodology and Applications - Proceedings of the European Safety and Reliability Conference, ESREL 2014*, 453–60. https://doi.org/10.1201/b17399-66.

Bouschery, Sebastian G., Vera Blazevic, and Frank T. Piller. 2023. "Augmenting Human Innovation Teams with Artificial Intelligence: Exploring Transformer-based Language Models." *Journal of Product Innovation Management* 40 (2): 139–53. https://doi.org/10.1111/jpim.12656.

Brachman, Michelle, Amina El-Ashry, Casey Dugan, and Werner Geyer. 2024. "How Knowledge Workers Use and Want to Use LLMs in an Enterprise Context." In *Extended Abstracts of the CHI Conference on Human Factors in Computing Systems*, 1–8. CHI '24. ACM. https://doi.org/10.1145/3613905.3650841.

Brauner, Philipp. 2024. "Mapping Acceptance: Micro Scenarios as a Dual-Perspective Approach for Assessing Public Opinion and Individual Differences in Technology Perception." *Frontiers in Psychology*. Frontiers Media SA. https://doi.org/10.3389/fpsyg.2024.1419564.

Brauner, Philipp, Manuela Dalibor, Matthias Jarke, Ike Kunze, István Koren, Gerhard Lakemeyer, Martin Liebenberg, et al. 2022. "A Computer Science Perspective on Digital Transformation in Production." *ACM Transactions on Internet of Things* 3 (2): 1–32. https://doi.org/10.1145/3502265.

Brauner, Philipp, Alexander Hick, Ralf Philipsen, and Martina Ziefle. 2023. "What does the public think about artificial intelligence? A criticality map to understand bias in the public perception of AI." *Frontiers in Computer Science* 5. https://doi.org/10.3389/fcomp.2023.1113903.

Brauner, Philipp, Ralf Philipsen, André Calero Valdez, and Martina Ziefle. 2019. "What happens when Decision Support Systems fail? — The Importance of Usability on Performance in Erroneous Systems." *Behaviour & Information Technology* 38 (12): 1225–42. https://doi.org/10.1080/0144929X.2019.1581258.

Brossard, Dominique, and Dietram A. Scheufele. 2013. "Science, New Media, and the Public." *Science* 339 (6115): 40–41. https://doi.org/10.1126/science.1232329.

Brynjolfsson, Erik, and Andrew McAfee. 2014. *The Second Machine Age: Work, Progress, and Prosperity in a Time of Brilliant Technologies*. New York.

Cath, Corinne. 2018. "Governing Artificial Intelligence: Ethical, Legal and Technical Opportunities and Challenges." *Philosophical Transactions of the Royal Society A: Mathematical, Physical and Engineering Sciences* 376 (2133): 20180080. https://doi.org/10.1098/rsta.2018.0080.

Cave, Stephen, Kate Coughlan, and Kanta Dihal. 2019. "'Scary Robots': Examining Public Responses to AI." In *Proceedings of the 2019 AAAI/ACM Conference on AI, Ethics, and Society*, 331–37. AIES '19. New York: ACM. https://doi.org/10.1145/3306618.3314232.

Chen, Lijia, Pingping Chen, and Zhijian Lin. 2020. "Artificial Intelligence in Education: A Review." *IEEE Access* 8: 75264–78. https://doi.org/10.1109/ACCESS.2020.2988510.

Chow, Gregory C. 1960. "Tests of Equality Between Sets of Coefficients in Two Linear Regressions." *Econometrica* 28 (3). https://doi.org/10.2307/1910133.

Connor, Melanie, and Michael Siegrist. 2010. "Factors Influencing People's Acceptance of Gene Technology: The Role of Knowledge, Health Expectations, Naturalness, and Social Trust." *Science Communication* 32 (4): 514–38. https://doi.org/10.1177/1075547009358919.

Crockett, Keeley, Matt Garratt, Annabel Latham, Edwin Colyer, and Sean Goltz. 2020. "Risk and Trust Perceptions of the Public of Artifical Intelligence Applications." In *2020 International Joint Conference on Neural Networks (IJCNN)*, 1–8. IEEE. https://doi.org/10.1109/ijcnn48605.2020.9207654.

Culkin, John M. 1967. "A Schoolman's Guide to Marshall McLuhan." *Saturday Review*, 51–53, 71–72.

Decker, Marie Christin, Laila Wegner, and Carmen Leicht-Scholten. 2024. "Procedural Fairness in Algorithmic Decision-Making: The Role of Public Engagement." *Ethics and Information Technology* 27 (1). https://doi.org/10.1007/s10676-024-09811-4.

Deng, Jia, Wei Dong, Richard Socher, Li-Jia Li, Kai Li, and Li Fei-Fei. 2009. "ImageNet: A Large-Scale Hierarchical Image Database." In *2009 IEEE Conference on Computer Vision and Pattern Recognition*, 248–55. https://doi.org/10.1109/CVPR.2009.5206848.

Diakopoulos, Nicholas. 2019. *Automating the News: How Algorithms Are Rewriting the Media*. Harvard University Press.

Efendić, Emir, Subramanya Prasad Chandrashekar, Cheong Shing Lee, Lok Yan Yeung, Min Ji Kim, Ching Yee Lee, and Gilad Feldman. 2021. "Risky Therefore Not Beneficial: Replication and Extension of Finucane Et Al.'s (2000) Affect Heuristic Experiment." *Social Psychological and Personality Science* 13 (7): 1173–84. https://doi.org/10.1177/19485506211056761.

Elena, Gianfranco, and Christopher W. Johnson. 2015. "Laypeople and Experts Risk Perception of Cloud Computing Services." arXiv. https://doi.org/10.48550/ARXIV.1509.06536.

Elsey, Jamie, and David Moss. 2023. "US Public Opinion of AI Policy and Risk." https://rethinkpriorities.org/research-area/us-public-opinion-of-ai-policy-and-risk/.

European Union. 2024. "Regulation (EU) 2024/1244 of the European Parliament and of the Council of 13 March 2024 laying down harmonised rules on artificial intelligence (Artificial Intelligence Act)." https://eur-lex.europa.eu/eli/reg/2024/1244/oj.

Fast, Ethan, and Eric Horvitz. 2017. "Long-Term Trends in the Public Perception of Artificial Intelligence." In *AAAI'17: Proceedings of the Thirty-First AAAI Conference on Artificial Intelligence*, 963–69.

Field, Andy. 2009. *Discovering Statistics Using SPSS*. 3. ed. Sage Publications.

Fischhoff, Baruch, Paul Slovic, Sarah Lichtenstein, Stephen Read, and Barbara Combs. 1978. "How Safe Is Safe Enough? A Psychometric Study of Attitudes Towards Technological Risks and Benefits." *Policy Sciences* 9 (2): 127–52. https://doi.org/10.1007/bf00143739.

Fox, John. 2016. *Applied Regression Analysis and Generalized Linear Models*. 3rd ed. Thousand Oaks, CA: SAGE Publications.

Friedman, Batya, and Helen Nissenbaum. 1996. "Bias in Computer Systems." *ACM Trans. Inf. Syst.* 14 (3): 330–47. https://doi.org/10.1145/230538.230561.

Fuchs, Christoph, and Adamantios Diamantopoulos. 2009. "Using Single-Item Measures for Construct Measurement in Management Research – Conceptual Issues and Application Guidelines." *Die Betriebswirtschaft* 9.

Gabriel, Iason. 2020. "Artificial Intelligence, Values, and Alignment." *Minds and Machines* 30 (3): 411–37. https://doi.org/10.1007/s11023-020-09539-2.

Gefen, David, and Kai Larsen. 2017. "Controlling for Lexical Closeness in Survey Research: A Demonstration on the Technology Acceptance Model." *Journal of the Association for Information Systems* 18 (10): 727–57. https://doi.org/10.17705/1jais.00469.

Geng, Mingmeng, Caixi Chen, Yanru Wu, Yao Wan, Pan Zhou, and Dongping Chen. 2025. "The Impact of Large Language Models in Academia: From Writing to Speaking." In *Findings of the Association for Computational Linguistics: ACL 2025*, 19303–19. Association for Computational Linguistics. https://doi.org/10.18653/v1/2025.findings-acl.987.

Hoffmann, Arvid O. I., Thomas Post, and Joost M. E. Pennings. 2015. "How Investor Perceptions Drive Actual Trading and Risk-Taking Behavior." *Journal of Behavioral Finance* 16 (1): 94–103. https://doi.org/10.1080/15427560.2015.1000332.

Holzinger, Andreas, Iztok Fister, Hans-Peter Kaul, and Senthold Asseng. 2024. "Human-Centered AI in Smart Farming: Toward Agriculture 5.0." *IEEE Access* 12: 62199–214. https://doi.org/10.1109/access.2024.3395532.

Hopfield, John J. 1982. "Neural Networks and Physical Systems with Emergent Collective Computational Abilities." *Proceedings of the National Academy of Sciences* 79 (8): 2554–58. https://doi.org/10.1073/pnas.79.8.2554.

Hristova, Tsvetelina, Liam Magee, and Karen Soldatic. 2024. "The Problem of Alignment." *AI & SOCIETY*. https://doi.org/10.1007/s00146-024-02039-2.

Hu, Krystal. 2023. "ChatGPT Sets Record for Fastest-Growing User Base." https://www.reuters.com/technology/chatgpt-sets-record-fastest-growing-user-base-analyst-note-2023-02-01 last accessed 2024-10-01.

Huang, Xingyu, Shanshan Dai, and Honggang Xu. 2020. "Predicting Tourists' Health Risk Preventative Behaviour and Travelling Satisfaction in Tibet: Combining the Theory of Planned Behaviour and Health Belief Model." *Tourism Management Perspectives* 33: 100589. https://doi.org/10.1016/j.tmp.2019.100589.

Ipsos. 2022. "Global Attitudes Towards Artificial Intelligence."

Jensen, Theodore, Mary Theofanos, Kristen Greene, Olivia Williams, Kurtis Goad, and Janet Bih Fofang. 2024. "Reflection of Its Creators: Qualitative Analysis of General Public and Expert Perceptions of Artificial Intelligence." *Proceedings of the AAAI/ACM Conference on AI, Ethics, and Society* 7: 647–58. https://doi.org/10.1609/aies.v7i1.31667.

Jungherr, Andreas. 2023. "Artificial Intelligence and Democracy: A Conceptual Framework." *Social Media + Society* 9 (3). https://doi.org/10.1177/20563051231186353.

Karaca, Ozan, S. Ayhan Çalışkan, and Kadir Demir. 2021. "Medical Artificial Intelligence Readiness Scale for Medical Students (MAIRS-MS) – Development, Validity and Reliability Study." *BMC Medical Education* 21 (1). https://doi.org/10.1186/s12909-021-02546-6.

Kelley, Patrick Gage, Yongwei Yang, Courtney Heldreth, Christopher Moessner, Aaron Sedley, Andreas Kramm, David T. Newman, and Allison Woodruff. 2021. "Exciting, Useful, Worrying, Futuristic: Public Perception of Artificial Intelligence in 8 Countries." In *Proceedings of the 2021 AAAI/ACM Conference on AI, Ethics, and Society*, 627–37. AIES '21. ACM. https://doi.org/10.1145/3461702.3462605.

Kim, Hee-Woong, Hock Chuan Chan, and Sumeet Gupta. 2007. "Value-Based Adoption of Mobile Internet: An Empirical Investigation." *Decision Support Systems* 43 (1): 111–26. https://doi.org/10.1016/j.dss.2005.05.009.

Krewski, Daniel, Michelle C. Turner, Louise Lemyre, and Jennifer E. C. Lee. 2012. "Expert Vs. Public Perception of Population Health Risks in Canada." *Journal of Risk Research* 15 (6): 601–25. https://doi.org/10.1080/13669877.2011.649297.

Lecun, Yann, Yoshua Bengio, and Geoffrey Hinton. 2015. "Deep Learning." *Nature* 521 (7553): 436–44. https://doi.org/10.1038/nature14539.

Lee, Sangwon, Myojung Chung, Nuri Kim, and S. Mo Jones-Jang. 2024. "Public Perceptions of ChatGPT: Exploring How Nonexperts Evaluate Its Risks and Benefits." *Technology, Mind, and Behavior* 5 (4). https://doi.org/10.1037/tmb0000140.

Leiner, Dominik Johannes. 2019. "Too Fast, Too Straight, Too Weird: Non-Reactive Indicators for Meaningless Data in Internet Surveys." *Survey Research Methods* 13 (3): 229–48. https://doi.org/10.18148/srm/2019.v13i3.7403.

Lerner, Jennifer S., Ye Li, Piercarlo Valdesolo, and Karim S. Kassam. 2015. "Emotion and Decision Making." *Annual Review of Psychology* 66 (1): 799–823. https://doi.org/10.1146/annurev-psych-010213-115043.

Lim, Sohye, and Hongjin Shim. 2022. "No Secrets Between the Two of Us: Privacy Concerns over Using AI Agents." *Cyberpsychology: Journal of Psychosocial Research on Cyberspace* 16 (4). https://doi.org/10.5817/cp2022-4-3.

Luhmann, Niklas. 1989. *Ecological Communication*. University of Chicago Press.

Makridakis, Spyros. 2017. "The Forthcoming Artificial Intelligence (AI) Revolution: Its Impact on Society and Firms." *Futures* 90: 46–60. https://doi.org/10.1016/j.futures.2017.03.006.

McCarthy, John, Marvin L. Minsky, Nathaniel Rochester, and Claude E. Shannon. 2006. "A Proposal for the Dartmouth Summer Research Project on Artificial Intelligence." *AI Magazine* 27 (4): 12–14.

McLuhan, Marshall. 1964. *Understanding Media: The Extensions of Man*. New York, NY: McGraw-Hill.

Moriniello, Flavio, Ana Martí-Testón, Adolfo Muñoz, Daniel Silva Jasaui, Luis Gracia, and J. Ernesto Solanes. 2024. "Exploring the Relationship Between the Coverage of AI in WIRED Magazine and Public Opinion Using Sentiment Analysis." *Applied Sciences* 14 (5): 1994. https://doi.org/10.3390/app14051994.

Müller, Vincent C., and Nick Bostrom. 2016. "Future Progress in Artificial Intelligence: A Survey of Expert Opinion." In *Fundamental Issues of Artificial Intelligence*, edited by Vincent C. Müller, 553–71. Springer.

Neri, Hugo, and Fabio Cozman. 2019. "The Role of Experts in the Public Perception of Risk of Artificial Intelligence." *AI & SOCIETY* 35 (3): 663–73. https://doi.org/10.1007/s00146-019-00924-9.

Neyer, Franz J., Juliane Felber, and Claudia Gebhardt. 2016. "Kurzskala Technikbereitschaft (TB, Technology Commitment)." *Zusammenstellung Sozialwissenschaftlicher Items Und Skalen (ZIS)*. https://doi.org/10.6102/ZIS244.

Niessen, Désirée, Constanze Beierlein, Beatrice Rammstedt, and Clemens M. Lechner. 2021. "Interpersonal Trust Short Scale (KUSIV3)." *ZIS – The Collection of Items and Scales for the Social Sciences*. https://doi.org/10.6102/ZIS292_EXZ.

Nissenbaum, H. 2001. "How Computer Systems Embody Values." *Computer* 34 (3): 120–19. https://doi.org/10.1109/2.910905.

Novozhilova, Ekaterina, Kate Mays, Sejin Paik, and James E. Katz. 2024. "More Capable, Less Benevolent: Trust Perceptions of AI Systems Across Societal Contexts." *Machine Learning and Knowledge Extraction* 6 (1): 342–66. https://doi.org/10.3390/make6010017.

Organisation for Economic Co-operation and Development. 2019. "OECD AI Principles." https://legalinstruments.oecd.org/en/instruments/OECD-LEGAL-0449.

Peacock, Walter Gillis, Samuel David Brody, and Wes Highfield. 2005. "Hurricane Risk Perceptions Among Florida's Single Family Homeowners." *Landscape and Urban Planning* 73 (2–3): 120–35. https://doi.org/10.1016/j.landurbplan.2004.11.004.

Peterson, Tarla, Markus Peterson, and William Grant. 2004. "Chapter Two: Social Practice and Biophysical Process." *The Environmental Communication Yearbook* 1: 15–32. https://doi.org/10.1207/s15567362ecy0101_2.

Pew Research Center. 2023. "What Americans Know about Everyday Uses of Artificial Intelligence." Pew Research Center. https://www.pewresearch.org/.

Pidgeon, Nick, and Baruch Fischhoff. 2011. "The Role of Social and Decision Sciences in Communicating Uncertain Climate Risks." *Nature Climate Change* 1 (1): 35–41. https://doi.org/10.1038/nclimate1080.

Puzanova, Z. V, A. G. Tertyshnikova, and U. O. Pavlova. n.d. "Technological Discourse in the Russian Media: Main Strategies for Representing Artificial Intelligence." *RUDN Journal of Sociology* 24: 747–63. https://doi.org/10.22363/2313-2272-2024-24-3-747-763.

Rammstedt, Beatrice, and Constanze Beierlein. 2014. "Can't We Make It Any Shorter? The Limits of Personality Assessment and Ways to Overcome Them." *Journal of Individual Differences* 35 (4): 212–20. https://doi.org/10.1027/1614-0001/a000141.

Recchia, Gabriel, Alexandra L. J. Freeman, and David Spiegelhalter. 2021. "How Well Did Experts and Laypeople Forecast the Size of the COVID-19 Pandemic?" Edited by Martial L. Ndeffo Mbah. *PLOS ONE* 16 (5): e0250935. https://doi.org/10.1371/journal.pone.0250935.

Robinson-Tay, Kathryn, and Wei Peng. 2024. "The Role of Knowledge and Trust in Developing Risk Perceptions of Autonomous Vehicles: A Moderated Mediation Model." *Journal of Risk Research*, 1–16. https://doi.org/10.1080/13669877.2024.2360923.

Rumelhart, David E., Geoffrey E. Hinton, and Ronald J. Williams. 1986. "Learning Representations by Back-Propagating Errors." *Nature* 323 (6088): 533–36. https://doi.org/10.1038/323533a0.

Russell, Stuart, Sabine Hauert, Russ Altman, and Veloso Manuela. 2015. "Robotics: Ethics of Artificial Intelligence." *Nature* 521 (7553): 415–18. https://doi.org/10.1038/521415a.

Sadek, Malak, Rafael A. Calvo, and Céline Mougenot. 2024. "Closing the Socio–Technical Gap in AI: The Need for Measuring Practitioners' Attitudes and Perceptions." *IEEE Technology and Society Magazine* 43 (2): 88–91. https://doi.org/10.1109/mts.2024.3392280.

Sanguinetti, Pablo, and Bella Palomo. 2024. "An Alien in the Newsroom: AI Anxiety in European and American Newspapers." *Social Sciences* 13 (11): 608. https://doi.org/10.3390/socsci13110608.

Sattar, Md. Abdus, and Kevin K. W. Cheung. 2019. "Tropical Cyclone Risk Perception and Risk Reduction Analysis for Coastal Bangladesh: Household and Expert Perspectives." *International Journal of Disaster Risk Reduction* 41: 101283. https://doi.org/10.1016/j.ijdrr.2019.101283.

Schwesig, Rebekka, Irina Brich, Jürgen Buder, Markus Huff, and Nadia Said. 2023. "Using Artificial Intelligence (AI)? Risk and Opportunity Perception of AI Predict People's Willingness to Use AI." *Journal of Risk Research* 26: 1053–84. https://doi.org/10.1080/13669877.2023.2249927.

Shneiderman, Ben. 2021. "Human-Centered AI." *Issues in Science and Technology* 37 (2): 56–61.

Siegrist, Michael, Carmen Keller, Hans Kastenholz, Silvia Frey, and Arnim Wiek. 2007. "Laypeople's and Experts' Perception of Nanotechnology Hazards." *Risk Analysis* 27 (1): 59–69. https://doi.org/10.1111/j.1539-6924.2006.00859.x.

Sindermann, Cornelia, Haibo Yang, Jon D. Elhai, Shixin Yang, Ling Quan, Mei Li, and Christian Montag. 2022. "Acceptance and Fear of Artificial Intelligence: Associations with Personality in a German and a Chinese Sample." *Discover Psychology* 2 (1). https://doi.org/10.1007/s44202-022-00020-y.

Sjöberg, Lennart. 1998. "Risk Perception: Experts and the Public." *European Psychologist* 3 (1): 1–12. https://doi.org/10.1027//1016-9040.3.1.1.

Slovic, Paul, Melissa L. Finucane, Ellen Peters, and Donald G. MacGregor. 2007. "The Affect Heuristic." *European Journal of Operational Research* 177 (3): 1333–52. https://doi.org/10.1016/j.ejor.2005.04.006.

Slovic, Paul, Baruch Fischhoff, and Sarah Lichtenstein. 1986. "The Psychometric Study of Risk Perception." In *Risk Evaluation and Management*, 3–24. Springer US. https://doi.org/10.1007/978-1-4613-2103-3_1.

Slovic, Paul, James Flynn, C. K. Mertz, Marc Poumadère, and Claire Mays. 2000. "Nuclear Power and the Public: A Comparative Study of Risk Perception in France and the United States." In *Cross-Cultural Risk Perception*, 55–102. Springer US. https://doi.org/10.1007/978-1-4757-4891-8_2.

Spiel, Katta, Oliver L. Haimson, and Danielle Lottridge. 2019. "How to Do Better with Gender on Surveys: A Guide for HCI Researchers." *Interactions* 26 (4): 62–65. https://doi.org/10.1145/3338283.

Statista. 2022. "Artificial Intelligence (AI) Worldwide – Statistics & Facts." https://www.statista.com/topics/3104/artificial-intelligence-ai-worldwide/dossier-chapter1 last accessed 28 Nov. 2022.

The Alan Turing Institute and Ada Lovelace Institute. 2023. "Understanding Public Attitudes to AI." The Alan Turing Institute. https://www.turing.ac.uk/.

Thomson, Mary E., Dilek Önkal, Ali Avcioğlu, and Paul Goodwin. 2004. "Aviation Risk Perception: A Comparison Between Experts and Novices." *Risk Analysis* 24 (6): 1585–95. https://doi.org/10.1111/j.0272-4332.2004.00552.x.

Thoreau, Henry David. 1854. *Walden; or, Life in the Woods*. Boston, MA: Ticknor; Fields.

Tourangeau, Roger, Lance J. Rips, and Kenneth Rasinski. 2000. *The Psychology of Survey Response*. Cambridge University Press. https://doi.org/10.1017/cbo9780511819322.

Verdurme, Annelies, and Jacques Viaene. 2003. "Consumer Beliefs and Attitude Towards Genetically Modified Food: Basis for Segmentation and Implications for Communication." *Agribusiness* 19 (1): 91–113. https://doi.org/10.1002/agr.10045.

Wang, Sai. 2025. "Public Perceptions of Artificial Intelligence in 20 Countries: Assessing Individual- and Country-Level Factors." *Cross-Cultural Research*. https://doi.org/10.1177/10693971251336803.

Wen, Chia-Ho Ryan, and Yi-Ning Katherine Chen. 2024. "Understanding Public Perceptions of Revolutionary Technology: The Role of Political Ideology, Knowledge, and News Consumption." *Journal of Science Communication* 23 (5). https://doi.org/10.22323/2.23050207.

Witte, Kim, and Mike Allen. 2000. "A Meta-Analysis of Fear Appeals: Implications for Effective Public Health Campaigns." *Health Education & Behavior* 27 (5): 591–615. https://doi.org/10.1177/109019810002700506.

Yigitcanlar, Tan, Kenan Degirmenci, and Tommi Inkinen. 2022. "Drivers Behind the Public Perception of Artificial Intelligence: Insights from Major Australian Cities." *AI & SOCIETY* 39 (3): 833–53. https://doi.org/10.1007/s00146-022-01566-0.